\documentclass{aa}  

\usepackage{graphicx}
\usepackage{txfonts}
\begin{document}

   \title{The aftermath of nova Cen 2013 (V1369 Cen)}

   \subtitle{%A binary still not in steady state five years after its nova outburst
   }

   \author{Elena Mason\inst{1}
          \and
          Steven N. Shore\inst{2}
          \and
          Jeremy Drake\inst{3}
          \and
          Steve B. Howell\inst{4}
          \and
          Paul Kuin\inst{5}
          \and
          Enza Magaudda\inst{6}
          }

   \institute{INAF-OATS, Via G.B. Tiepolo 11, 34143 Trieste, Italy\\
              \email{elena.mason@inaf.it}
         \and
          Dipartimento di Fisica "Enrico Fermi", Universit\'a di Pisa \& INFN-Pisa, Largo Pontecorvo 3 , 56127 Pisa, Italy 
             \and
        Smithsonian Astrophysical Observatory, MS-03, 60 Garden Street, Cambridge MA 02138, USA
        \and
        Space Science and Astrobiology Division, NASA Ames Research Center, M/S 245-6, Moffett Field, CA 94035, USA
        \and
        Mullard Space Science Laboratory, University College London, Holmbury St Mary, Dorking, Surrey RH5 6NT, UK
        \and
        Institut f\"ur Astronomie \& Astrophysik, Eberhard Karls Universit\"at T\"ubingen, Geschwister-Scholl-Platz, 72074 T\"ubingen, Germany
        }

   \date{}

% \abstract{}{}{}{}{} 
% 5 {} token are mandatory
 
  \abstract
   {Classical nova progenitors are cataclysmic variables and very old novae are observed to match high mass transfer rate and (relatively) long orbital period systems. However, the aftermath of a classical nova has never been studied in detail. }
   {To probe the aftermath of a classical nova explosion in cataclysmic variables and observe as the binary system relaxes to quiescence.}
  {We used multi-wavelength time resolved optical and near-infrared spectroscopy for a bright, well studied classical nova five years after outburst. We were able to disentangle the contribution of the ejecta at this late epoch using its previous characterization, separating  the ejecta emission from that of the binary system.}
   %{We observed through multi-band time resolved spectroscopy a well studied classical nova five years after outburst. The previous ejecta characterization allowed disentangling the ejecta emission from that of the binary system.}
   {We determined the binary orbital period ($P$=3.76 hr), the system separation and mass ratio ($q\gtrsim$0.17 for an assumed white dwarf mass of 1.2 M$_\odot$). We find evidence of an irradiated secondary star and no unambiguous signature of an accretion disk, although we identify a second emission line source tied to the white dwarf with an impact point. The data are consistent with a bloated white dwarf envelope and the presence of unsettled gas within the white dwarf Roche lobe. }
   {At more than 5 years after eruption, it appears that this classical nova has not yet relaxed. }

   \keywords{classical novae --
                cataclysmic variables --
                star: nova Cen 2013
               }

   \maketitle
%
%-------------------------------------------------------------------

\section{Introduction}
The classical nova (CN) V1369 Cen (= nova Cen 2013 hereafter abbreviated as Cen) was found in outburst on  2013 Dec  2.692UT (Waagen 2013). The event has been studied from maximum to late nebular phase with high resolution optical and UV spectroscopy, characterizing the ejecta dynamics, geometry and density (Mason et al. 2018). 
More than five years after outburst we observed the object again, with the intent of catching the underlying binary in its early relaxation stage and determining the timescale for the system to return to quiescence. 
Evidence exists of an early accretion disk formation for only a few CN.  For example, photometry of nova Her 1991 (Leibowitz et al. 1992, Leibowitz 1993), U Sco (Schaefer et al. 2011, Ness et al. 2012), and nova LMC 1968 (Kuin et al. 2020) showed broad band eclipses while still declining from the maximum light (i.e. at $\geqslant$3-4 mag from peak brightness). Thoroghgood et al. (2001) detected accretion disk double peaked emission lines in U~Sco $\sim$50 days after outburst. 

In this paper we present results from the analysis of optical and NIR time resolved spectroscopy obtained for Cen during two full non-contiguous nights in April 2019, i.e. 5.4 years after outburst.  We also present contemporaneous {\em Swift} UVOT photometry and {\em Chandra} spectroscopy.
   
%--------------------------------------------------------------------

\section{Observations and data reduction}
\subsection{VLT spectroscopic observation}
Observations were performed in Paranal at the Very Large Telescope, UT2, equipped with XShooter,  a three armed (UVB, VIS, and NIR) spectrograph  (Vernet et al. 2011), covering the wavelength range 3000~\AA \ -- 2.4~$\mu$m with the slit width dependent resolution 5000$<$R$<$9000 (see Table~\ref{table:xsh_obs} for the log of observations). 

Novae have orbital periods in the range 1.5 hr to a few days, the period distribution peaking at about 3-4 hr (e.g. Fuentes-Morales et al. 2021). To secure the period of Cen, detect possible intra- and trans- orbit variations, as well as short and mid term variability, it was necessary to observe for two full nights separated by about two weeks. Each night was sampled with an exposure time of about 400 sec to secure a cadence of 10 min, including overheads. 
Telescope nodding was preferred to staring mode observations because of verified better  pipeline performances in the flux calibration (especially in the blue part of the UVB arm and in the K band), and possibly to better subtract the sky in the NIR arm. 

Spectrophotometric standards were observed, with the same instrument setup, only at the beginning and the end of the night to not interrupt the time resolved spectroscopy sequence on Cen. The seeing was unfortunately not consistent through the night, which led to uncertainty in the flux calibration due to variation in the fraction of the light entering the 
spectrograph slit. 
To  limit differential color losses we used the same guide star for the whole night during both nights. In addition, the science target was reacquired every 1 to 1.5 hr (depending on the distance from meridian passage) to reset the slit orientation at the parallactic angle. The Atmospheric Dispersion Compensators (ADC) mounted in the UVB and VIS arms, further limited the color losses for zenith distances $<$60$^o$.
Telluric spectra were obtained at the beginning and at the end of the night. 
 
The data were reduced with the ESO XShooter pipeline (Modigliani et al. 2010) version 3.2.0 with {\it Gasgano} (v. 2.4.8, Hanuschik and Amico 2000) and {\it esorex}. The {\it esoreflex} version of the pipeline was too demanding in terms of CPU and not very user-friendly to be considered. 
We used the "offset" recipe since the "nod" procedure (i.e. the recipe appropriate for the adopted observing strategy) stacks all the exposures taken within an observing block. However, we also used the "stare" recipe for the UVB arm exposures, especially if affected by poor seeing. The "stare" and the "offset" recipes, because of the different background subtraction procedures, show differences of about 2\% in the flux calibrated spectra taken with decent seeing. However, they produce different emission line profiles in the case of spectra taken with poor seeing conditions. 

Flux calibration was performed only with the standard star observed at the end of the night (EG~274) since the standard at the beginning of the night (LTT~4816) has an incomplete flux table for the NIR wavelengths. EG~274 was used together with the observatory standard of the month (LTT~3218) to assess the accuracy of the flux calibration. The observatory standard of the night, however, was observed with a 5" slit. 

We decided to not remove the telluric absorptions using either the telluric star or the ESO dedicated software ({\it molecfit}, Smette et al. 2015) since, working with line profiles (rather than line fluxes), we prefer to see where the information is missing rather than risk tampering with the true emission line structure.   
For the same reason, we used  ESO {\it skycalc}\footnote{https://www.eso.org/observing/etc/bin/gen/form?INS.MODE=swspe ctr+INS.NAME=SKYCALC} to generate a list of NIR sky emission lines which, if not properly removed during the reduction process, were simply removed by connecting the median values of a few bounding pixels. 

\subsection{{\em Swift} photometry}
We attempted {\em Swift} UV spectroscopy and photometry  contemporary to the VLT observations, thus to better constrain the Cen spectral energy distribution (SED). Unfortunately, the UVOT slitless spectra were contaminated by nearby stars so  we are not sure the extracted spectral slope is reliable. The spectra did not show any significant emission line. Hence, we rely only on the photometry for any flux estimate. This was obtained on Apr 21 2019 (i.e. in between the two VLT runs) and two months later. The log of the UVOT observations is in Table~\ref{table:log_uv} together with the flux measures (see also Sec.~\ref{components}). 

The photometry was reduced using the {\em Swift} CALDB (Breeveld et al. 2011) and the {\em Swift} {\tt uvotsource} tool. Standard aperture and the revised sensitivity loss tables were adopted in the data reduction. 

\subsection{Chandra x-ray observation}
   \begin{figure}
   \centering
   \includegraphics[angle=0,width=9cm]{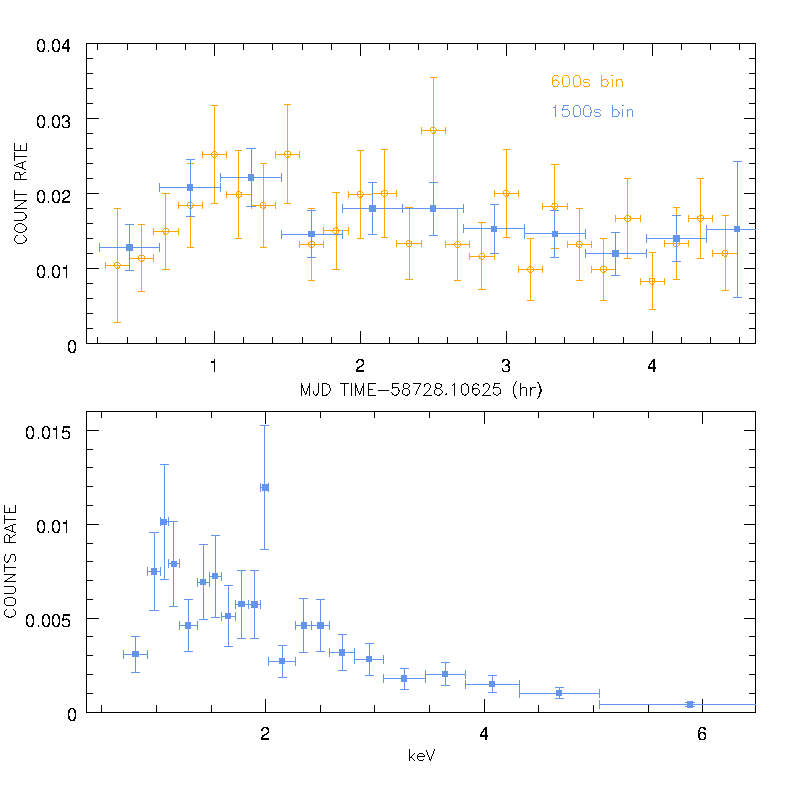}
      \caption{Top: the {\em Chandra} light curve of Cen in bin of 600 and 1500 seconds.  Bottom: the ACIS-S spectrum of Cen. 
              }
         \label{xplots}
   \end{figure}
%-----------------------------------------------------------------
We used {\it Chandra} to further constrain the interpretation of the source, and to search for any X-ray periodicity that could be the signature of a magnetic white dwarf (WD). We observed Cen with the low resolution ACIS-S spectrograph on board {\em Chandra} for $\gtrsim$4 hr. The journal of observations is in Table~\ref{table:x_obs}. The data reduction was performed using the CIAO software\footnote{\url{http://cxc.harvard.edu/ciao/}}, in particular, its  \textit{chandra\_repro} pipeline for the re-processing, and the packages \textit{specextract} and \textit{dmextract}, to extract the Cen spectrum and light curve, respectively. We verified that using different parameters in the extraction process (e.g. detection threshold and background annulus size) did not significantly affect the results. Fig.\ref{xplots} shows the light curve (in 600 and 1500 s time bins, top), and the spectrum  (bottom). 
After preliminary model fitting, which did not help constrain the exact energy distribution of the x-ray emitting source, we determined the absolute flux calibration of the spectrum  independent of any model assumption, relying instead on the collected photon energy. The absolute flux calibration was obtained using the instrument response matrix, i.e. the effective area in cm$^2$. In so doing we verified that the signal below $\sim$7~\AA\ and above 12.4~\AA\ (i.e. $\sim$1.8 to 1.0 keV) is too uncertain to be considered. Therefore, when computing the integrated flux and total luminosity we limited ourselves to that wavelength range (Section \ref{components}). 

% One column figure
%----------------------------------------------------------------- 
   \begin{figure}
   \centering
   \includegraphics[angle=90,width=9cm]{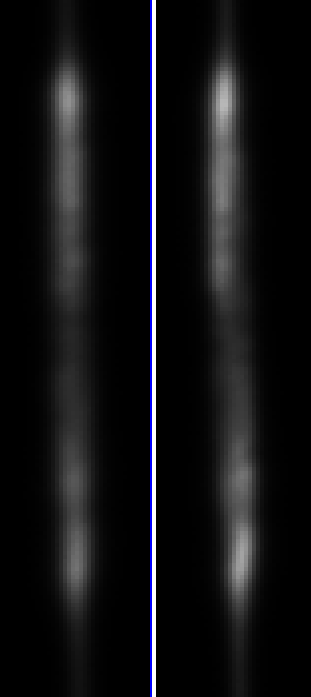}
   \includegraphics[width=9cm]{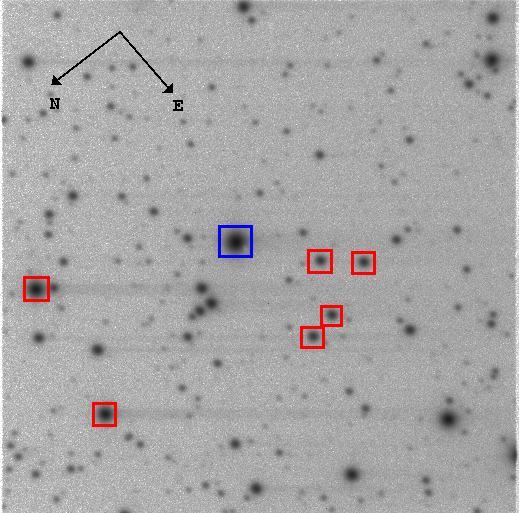}
      \caption{Top: Examples of the [O~III]$\lambda$4959 emission line as observed in the 2D raw frames. The bright spots correspond to the line structures/peaks in the Fig.\ref{lineProfile} profile. The wavelength increases from left to right. Bottom: The XShooter acquisition frame. The blue square encloses Cen, while the red ones cover the comparison stars that were used to estimate the seeing PSF. The field shown is 1.47 arcmin squared. 
              }
         \label{acq}
   \end{figure}
%-----------------------------------------------------------------
\section{Data analysis and results}\label{results}
%----------------------------------------------------------------- 
   \begin{figure} 
   \centering
   \includegraphics[width=9cm]{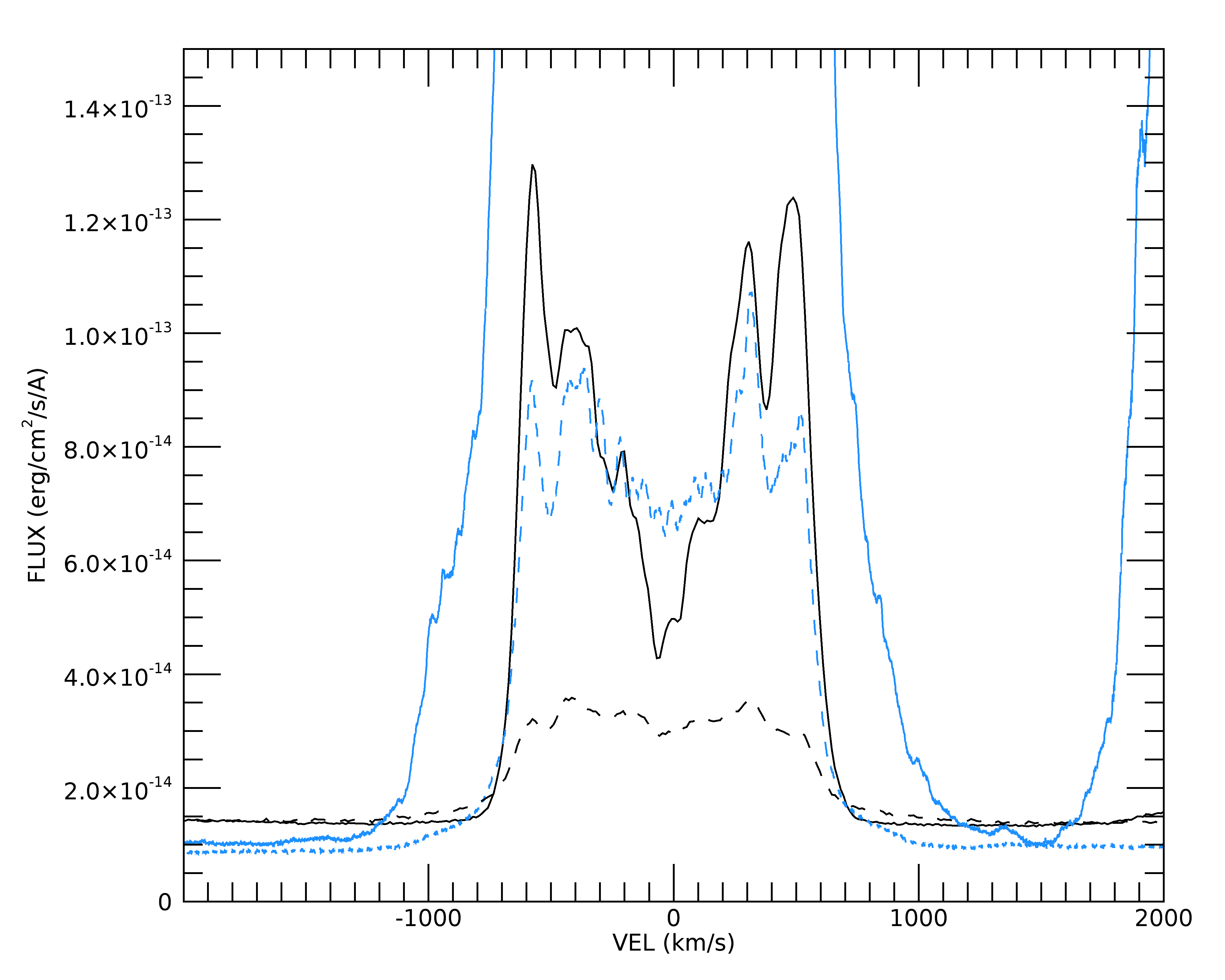}
      \caption{[[O~III]~$\lambda$4959 and H$\beta$ line profiles in April 2019 (black solid and dashed lines, respectively), together with the same line profiles from March 2016 (blue  lines) showing the change of the emission measure per velocity bin. The [O~III] emission lines were about 10 times stronger in 2016 than 2019.
              }
         \label{lineProfile}
   \end{figure}
%-----------------------------------------------------------------

\subsection{Acquisition images and 2D frames}\label{sec:2d}
During the XShooter data reduction process, inspection of the raw 2D frames revealed that the [O~III]$\lambda$4959 showed emission lines structures (blobs) that are tilted in some exposures (see Fig.\ref{acq}, top). This effect is evident only in the [O~III]$\lambda$4959 line since it is the strongest non-saturated line in the XShooter UVB+VIS+NIR spectrum\footnote{The [O III]$\lambda$5007 emission line is saturated in all the exposures except the shortest.  The [N~II]+H$\alpha$ blend is present in order 25 and order 24 of the VIS arm. It is saturated in order 25 and much weaker in order 24.}.
The presence of tilted emission structures suggests that the ejecta were possibly resolved. To check this hypothesis we measured the Point Spread Function (PSF) of Cen and 5 nearby stars in each acquisition frame (i.e. seven frames in night 1, and eight in night 2; Fig.\ref{acq}, bottom). A Student t-test showed that Cen is somewhat larger than the average PSF in 47\% of the frames. Hence, we conclude that the ejecta are marginally resolved.
This would explain the occasional variation of the [O~III] line profile observed in some spectra, especially on the second night. Possibly depending on the seeing variation and photon statistic, the centering of Cen in the slit might have been affected by bright structures in the ejecta (we used the Johnson V filter for the target acquisition, which at 5000 \AA\ transmits almost 80\% of the incoming light -- see the instrument manual). 
Partial resolution of the ejecta may also explain the large uncertainties in the Gaia DR2 distance (317$^{+819}_{-130}$ pc, Bailer-Jones et al. 2018).

We can use the resolved images to constrain the distance to Cen from its nebular parallax. Using the acquisition frames where the Cen PSF was resolved, we estimate the average size of the nebula to be $<$FWHM$>$=0.44$\pm$0.11 arcsec, where the individual values were calculated from $FWHM=\sqrt{FWHM_{Cen}^2-\overline{FWHM}_{field~stars}^2}$.
The expansion velocity to adopt in the nebular parallax computation is not that observed during maximum or early decline but that of the currently emitting ejecta. At the time of the XShooter observations, the nebular emission lines are produced mainly by the inner ejecta because of the reduced emission measure of the highest velocity gas. Fig.\ref{lineProfile} shows that the strongest ejecta structures extend up to $\sim$800 km/s along the line of sight. For an ejecta inclination of 40 deg (Mason et al. 2018) this translates to an absolute maximum velocity of $\sim$1200 km/s and to a projected velocity  on the sky plane of $\sim$950 km/s. The latter, produces a distance of about 2.4$^{+0.9}_{-0.5}$ kpc, in agreement with our previous determination given the uncertainties ($\simeq$2 kpc, Mason et al. 2018). 

   \begin{figure*}
   \centering
   \includegraphics[width=18cm,angle=0]{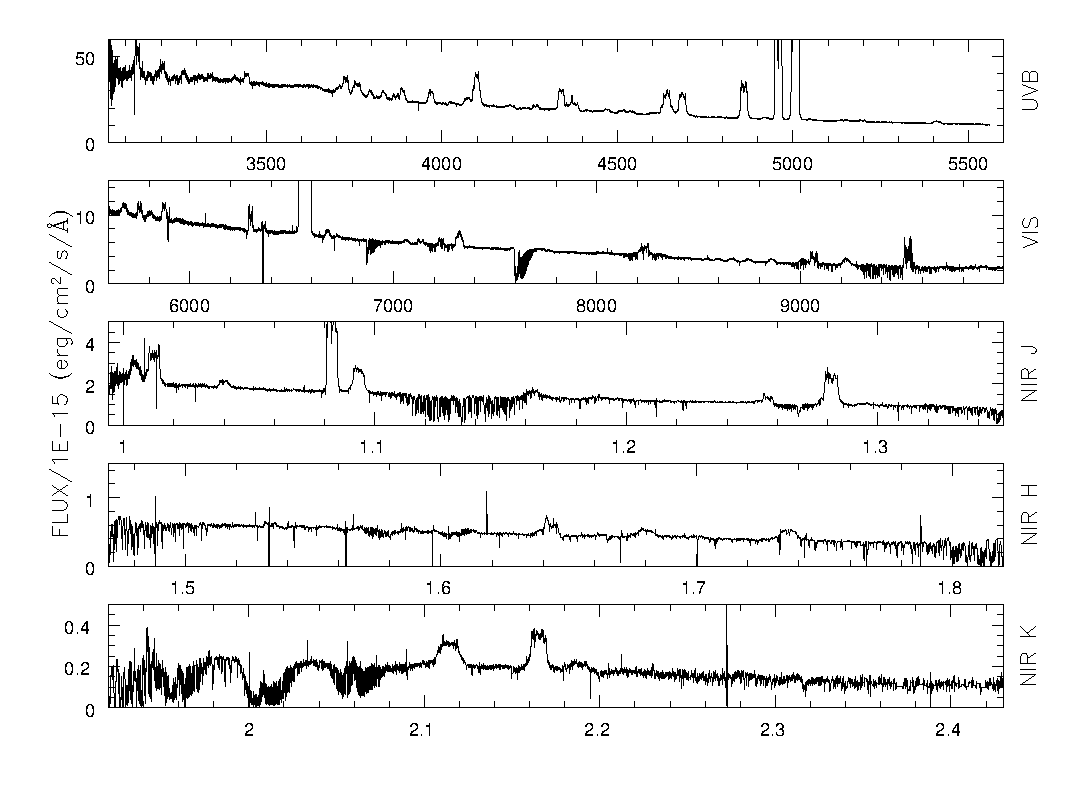}
      \caption{Cen fiducial XShooter spectrum. UVB and VIS arm (top two panels) x-axes are in unit of \AA; NIR arm (bottom 3 panels) x-axes units are in $\mu$m. The spectrum is not corrected for reddening.  
              }
         \label{spc4all}
   \end{figure*}

\subsection{The emission-line spectrum}\label{sec:emlinespc}
\begin{figure*}[h!]
   \centering
   \includegraphics[width=5.5cm]{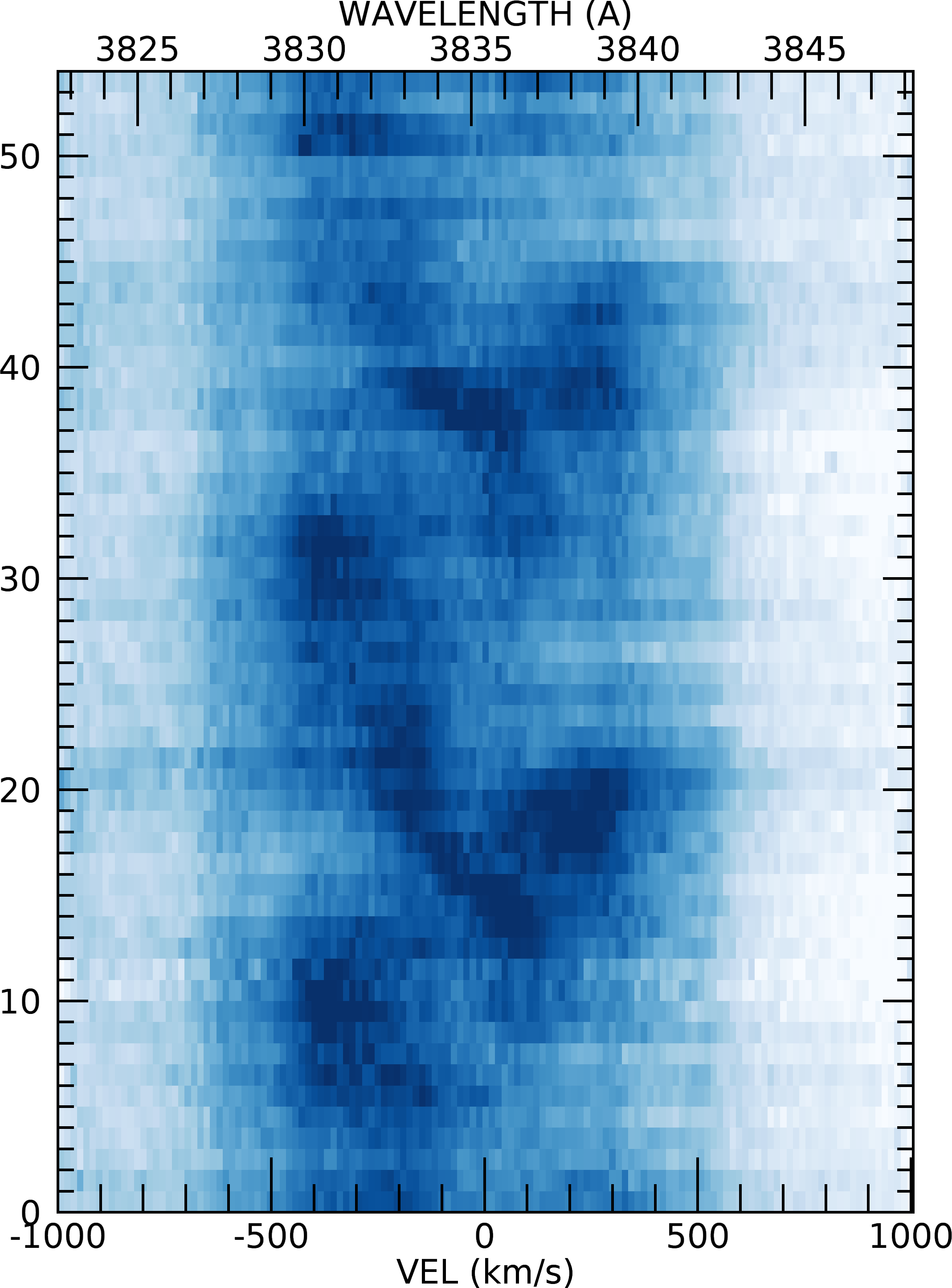}
   \includegraphics[width=5.5cm]{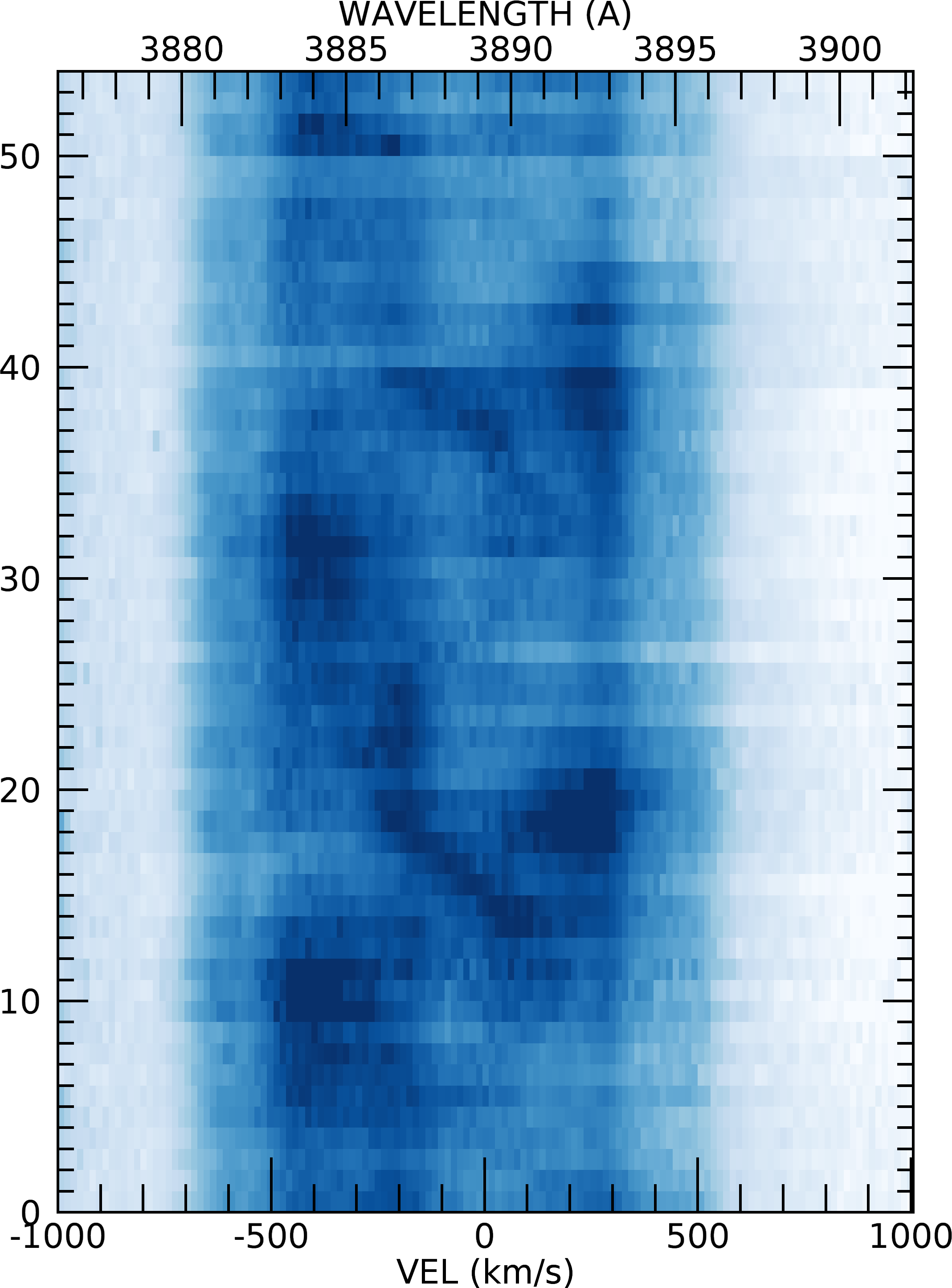}
   \includegraphics[width=5.5cm]{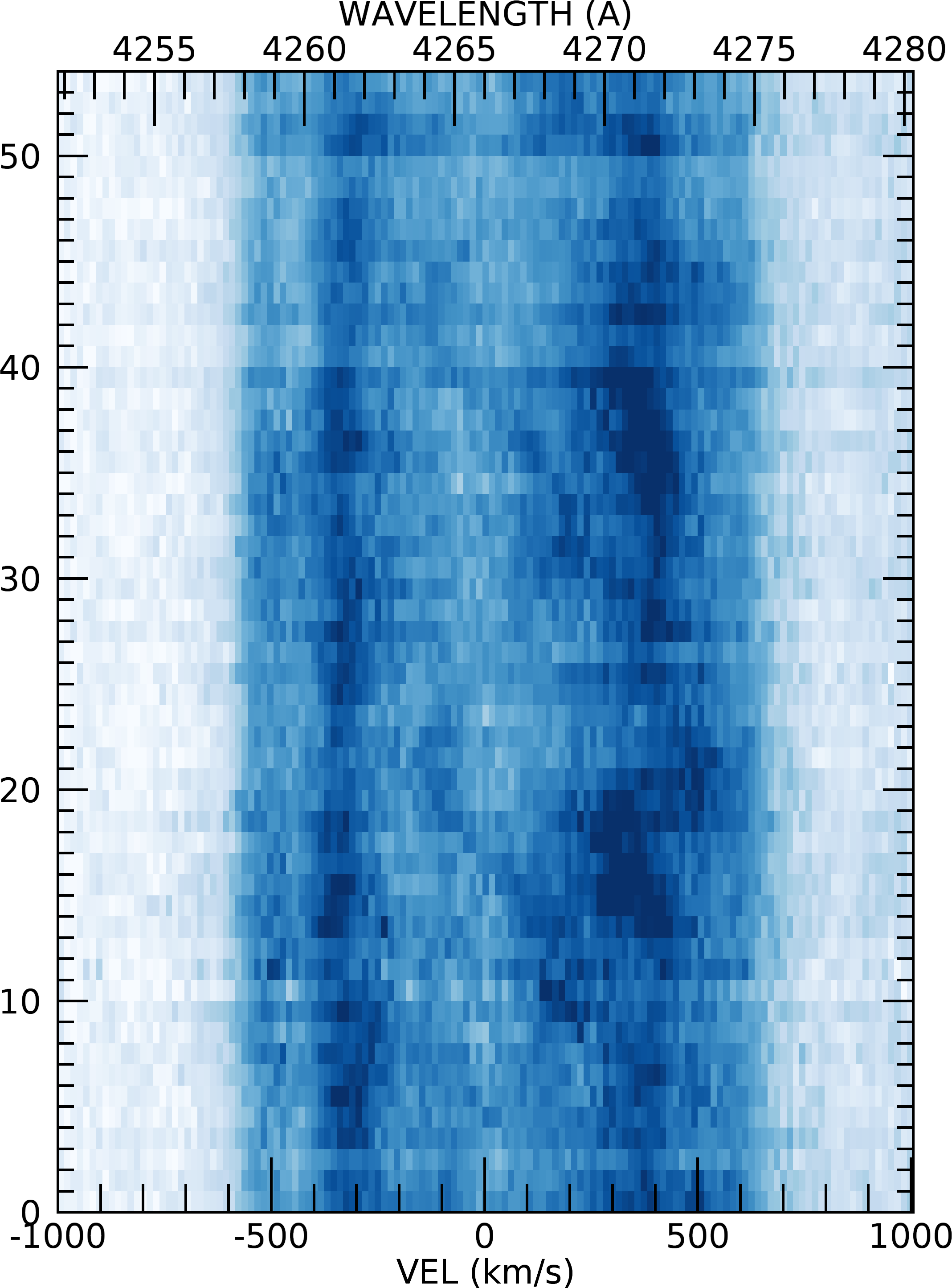}
   \includegraphics[width=5.5cm]{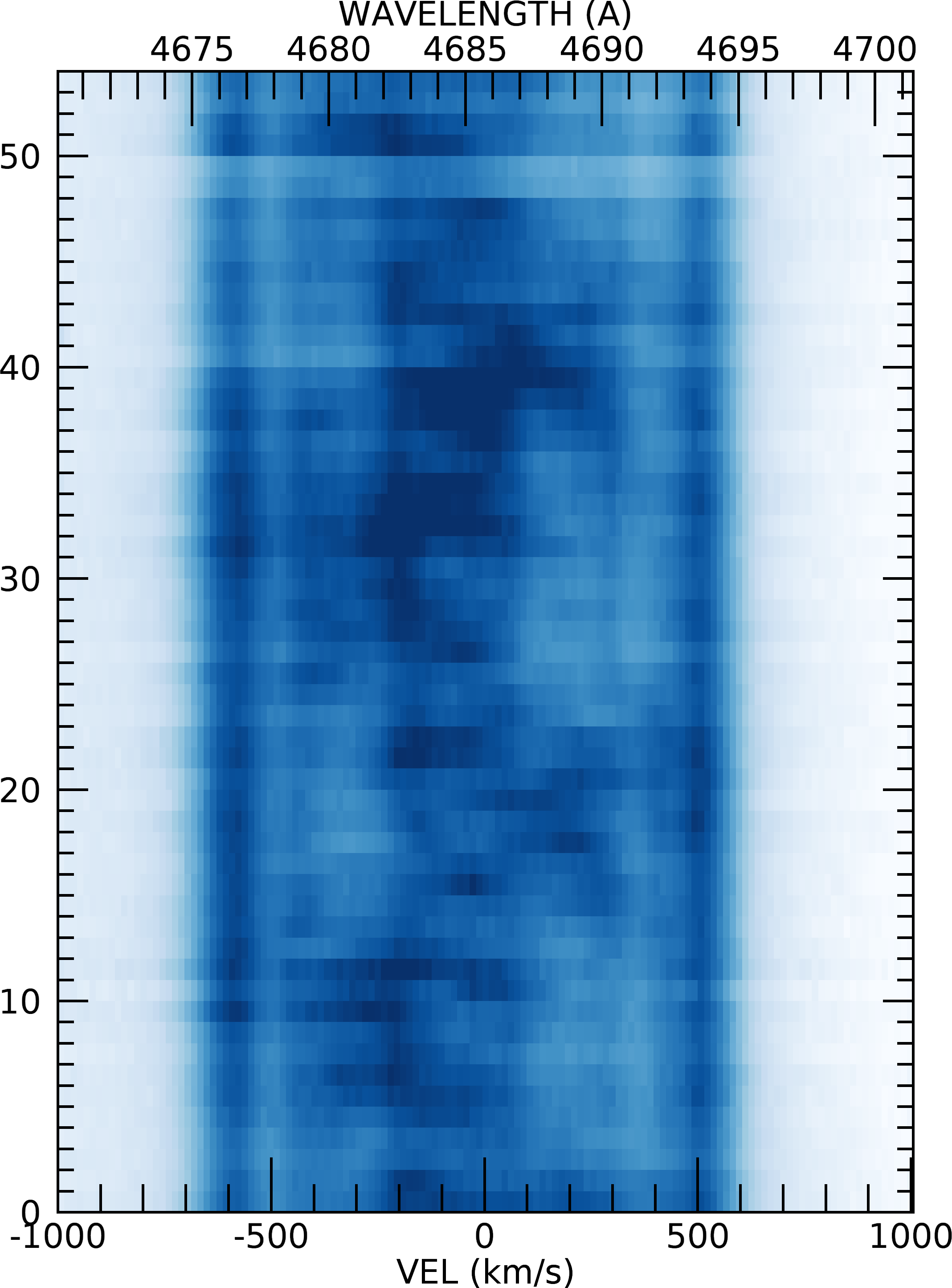}
   \includegraphics[width=5.5cm]{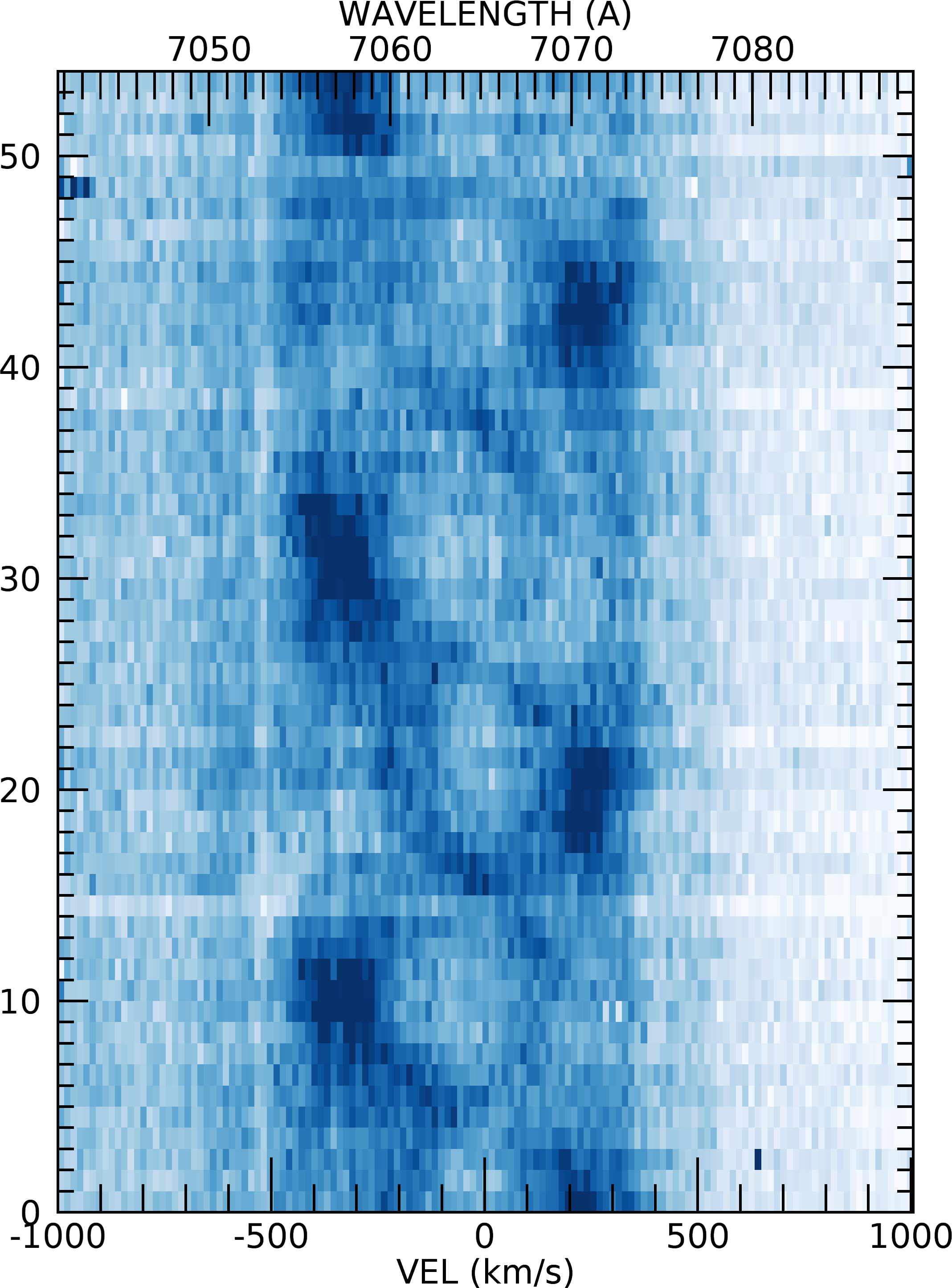}
   \includegraphics[width=5.5cm]{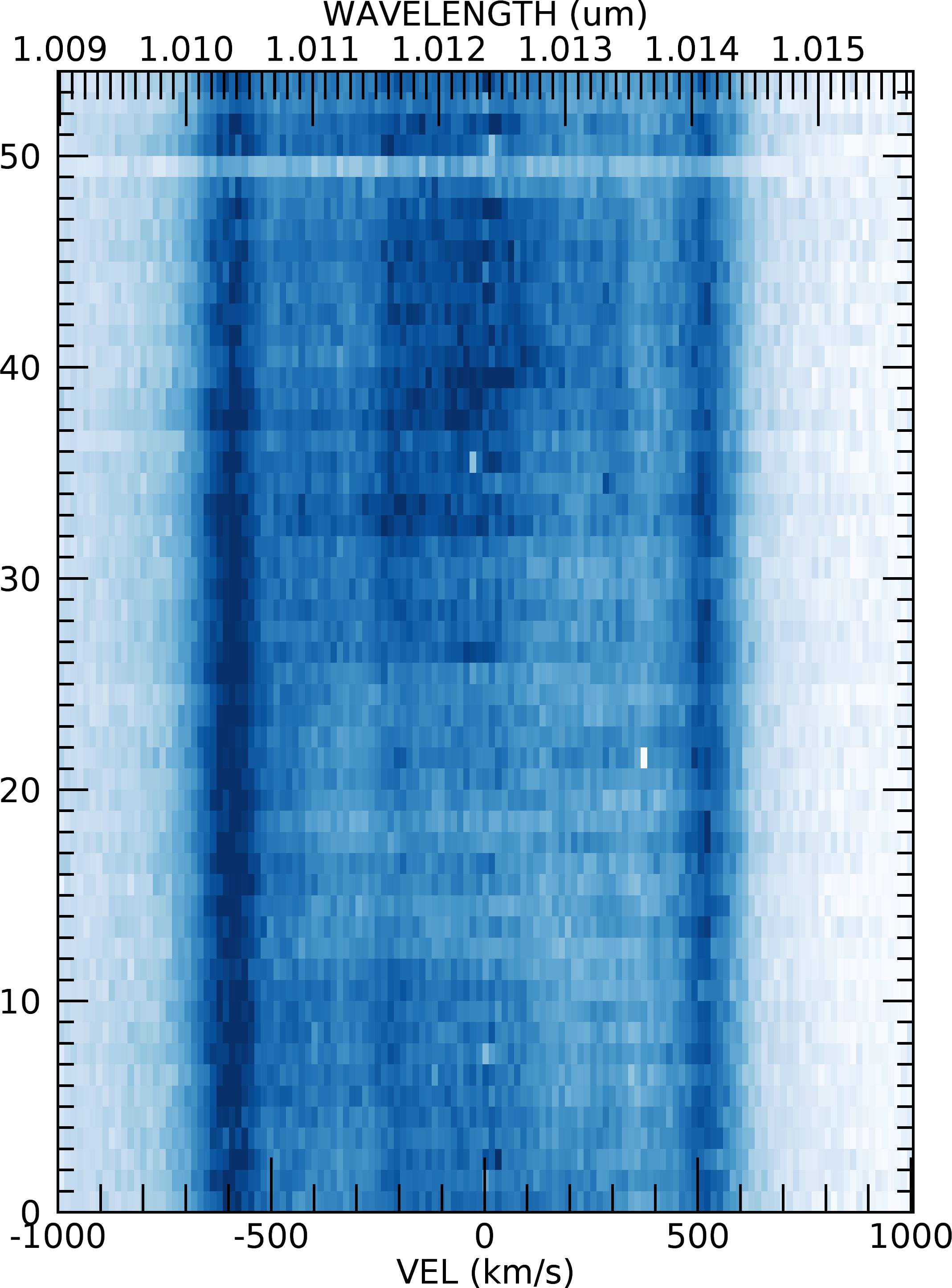}
   \includegraphics[width=5.5cm]{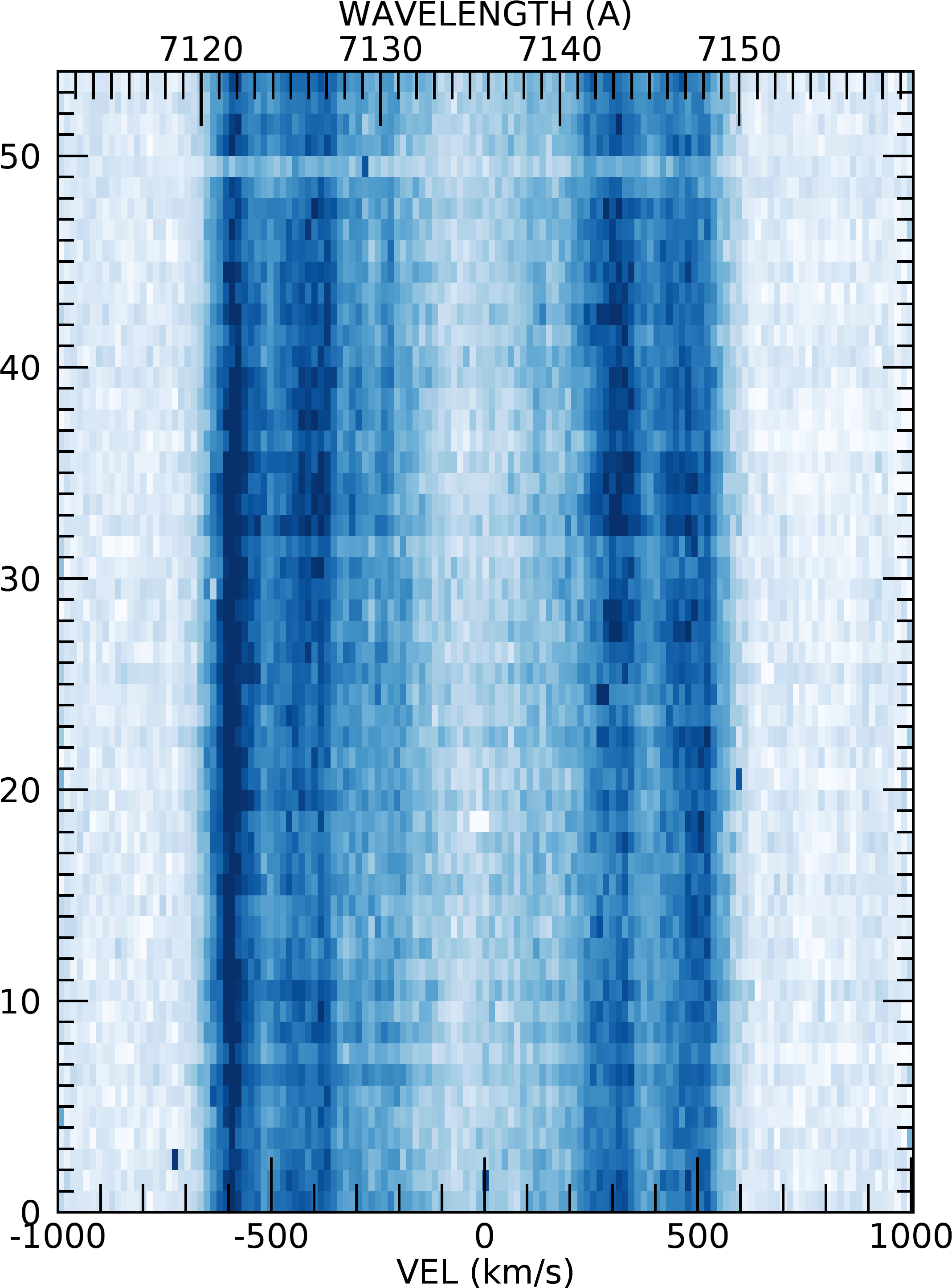}
   \includegraphics[width=5.5cm]{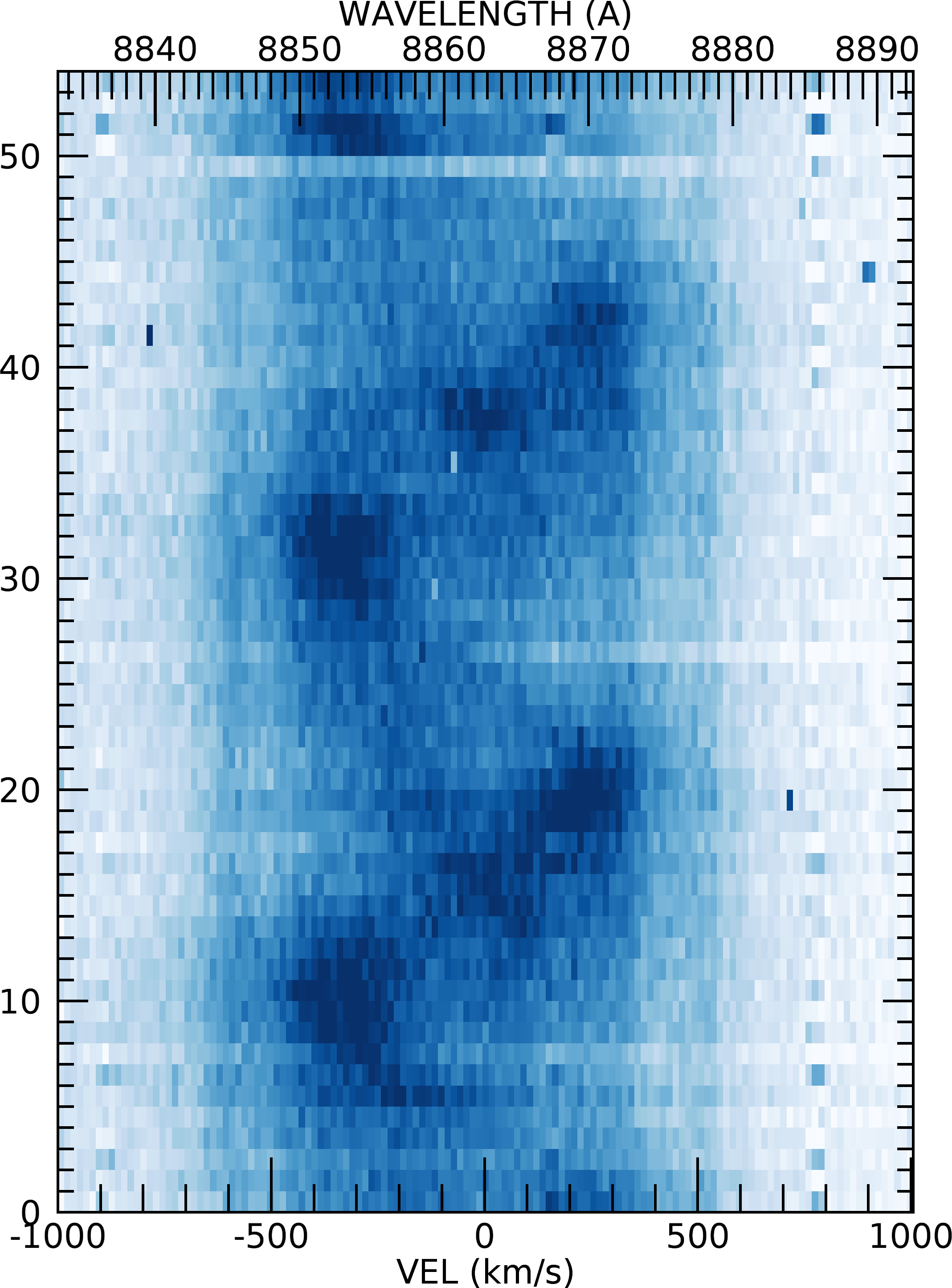}
   \includegraphics[width=5.5cm]{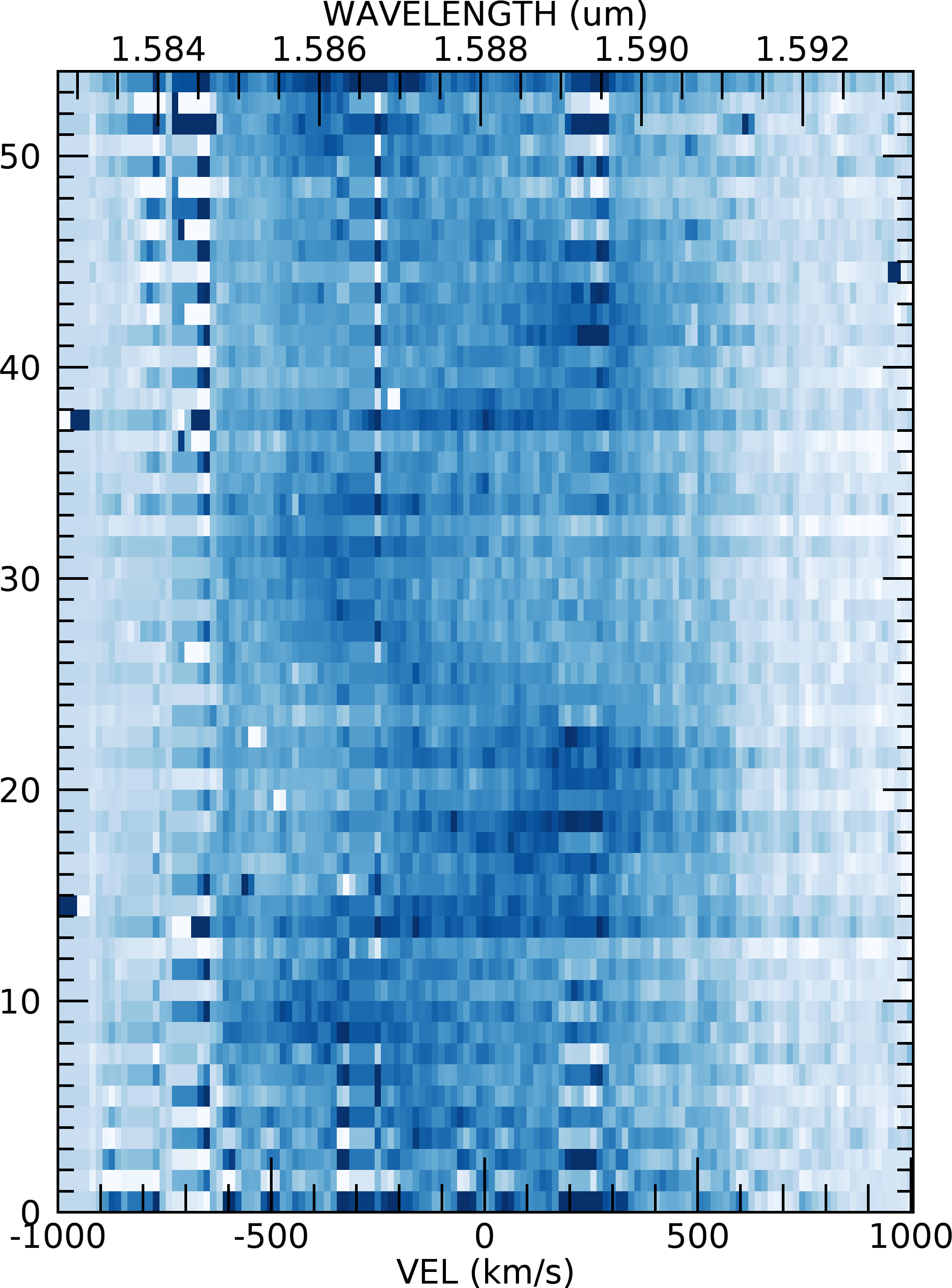}
   \caption{Selected trailed spectrograms from night 1. Darker regions indicates stronger emission (linear intensity scale). From top to bottom and left to right: H$\eta$, H$\zeta$, C~II~$\lambda$4266, He~II~$\lambda$4686, He~I$\lambda$7065, He~II~$\lambda$10123, [Ar~III] $\lambda$7135, H~Paschen $\lambda$8862, and H~Brackett $\lambda$15880. The y-axis in each panel indicates the spectrum number in the night time series. The fuzzballs are particularly evident at maximum blueshift around spectra number 10 and 30, and at maximum redshift around spectra $\sim$20 and 40 in the H$\eta$, H$\zeta$, He~I$\lambda$7065, and H~Paschen $\lambda$8862 lines. }
              \label{trails}%
    \end{figure*}

Figure~\ref{spc4all} shows the emission lines dominating Cen spectrum. 
We identify over 80 emission lines in the UVB+VIS+NIR spectrum of Cen. Of those, several have a purely ejecta origin, while the others are a varying mix of ejecta and binary contributions (see Appendix~\ref{sec:linelist}). Pure ejecta lines are identified by their invariant profile and comprise forbidden transitions. Numerous permitted transitions also reveal a contribution from the binary underneath by their variable profile. The observed variations are minimal but become obvious when trailed spectrograms are produced. The narrow s-wave in the transitions of the higher Balmer and Paschen H series, He I, and C~II$\lambda$4267 are clearly of binary origin. Similarly, of binary origin are features we nicknamed "fuzzballs" that appear almost in anti-phase with the narrow s-wave. The fuzzballs are most visible in the higher Balmer, Paschen, and Brackett series lines and the He~I transitions. In addition, in the He~I and in the NIR H lines they morph into a complete sinusoidal curve (the fuzzball-wave).
Fig.\ref{trails} shows a selection of trailed spectrograms  of nebular and permitted transitions. The visibility of the binary s-wave and fuzzballs in the H lines indicates that they have higher opacity and a flatter decrement than the ejecta (which satisfy case-B optically thin conditions). The presence of the fuzzball-wave in the H Paschen and Brackett lines where the s-wave disappears indicates higher opacity of the gas producing the fuzzballs than that producing the s-wave.

\subsubsection{The radial velocity curve and the binary parameters}\label{rvmass}
The contrast between the binary and the ejecta emissions is low and does not improve significantly after subtracting the ejecta template (e.g. from spectra of previous epochs or from a purely nebular emission line profile). Hence, the binary emissions cannot be fit or measured in individual spectra, but their position and motion can be measured by cursor position in the trailed spectrograms, especially for the s-wave. We did so for the s-wave in  H~I $\lambda$3797, 3835, 3889 and 3970, He~I $\lambda$4026, 4471\footnote{This line, however, seems biased by a blending emission.}, 5876, 6678 and 7065, and C~II $\lambda$4267. We measured also the fuzzball-wave for the He~I $\lambda$6678 and 7065 trailed spectrograms . 

\begin{figure*}
\centering
\includegraphics[width=18cm]{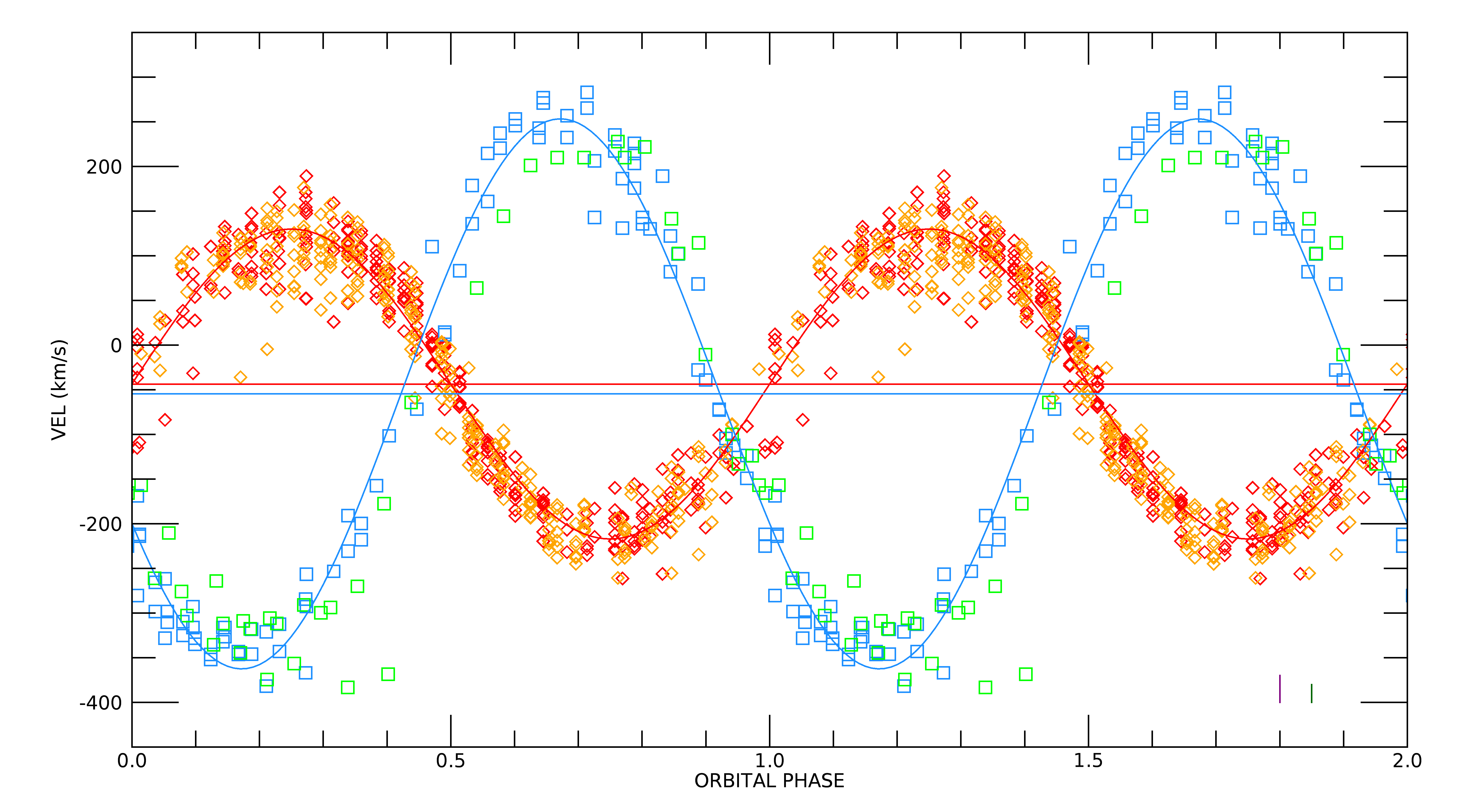}
\caption{Radial velocities of the s-wave (red and orange diamonds for night 1 and night 2, respectively), and of the fuzzball-wave (blue and green squares for night 1 and night 2, respectively). The red sine curve and horizontal lines are the night 1 best fit radial and $\gamma$ velocity for the s-wave, respectively. The blue sine curve and horizontal line are the night 1 best fit radial and $\gamma$ velocity, respectively, for the fuzzball emission. See Section~\ref{fuzz} for details. The two vertical bars on the bottom right corner of the panel represent the radial velocity measurements 1$\sigma$ uncertainties: 30 km/s in purple color for the UVB arm measurements and 20 km/s in dark green for the VIS arm measurements. }
\label{rv}%
 \end{figure*}

We fit the  Heliocentric radial velocity curve of each line separately, each ion (i.e. H~I, He~I, and C~II) separately, and finally all measures together using 
\begin{equation}
    v_r=\gamma+K\times\sin{[(t-t_0)/P]}
\end{equation}
where $\gamma$ is the systemic velocity, $K$ is the amplitude of the s-wave and can be negative, $P$ is the period and $t$ the mid-MJD of each exposure. We chose the time of the first red-to-blue crossing in each night  for $t_0$. The best fit parameters, $\gamma$, $K$, and $P$, and their associated errors were determined through a  bootstrapping technique (e.g. Wall and Jenkins 2012). Before combining the radial velocity measures from different transitions we always checked that the $\chi^2$ distribution of the individual fits were comparable. We also overplotted the measures from a given line on the trailed spectrogram of other transitions to visually check consistency. 
Table~\ref{rvsfit} gives the best fit for the s-wave parameters obtained by combining all the H~I, He~I and C~II measurements for each night. It also gives the best fit obtained for individual atomic species or ions. Figure~\ref{rv} shows the H and He~I radial velocity measures together with the global best fit sine functions from night 1.

The set of parameters are fairly similar between the two nights. 
The time elapsed between the s-wave curve of the two nights is 12.995$\pm$0.002 days (i.e. shifting night 1 by 12.995 days, its data points superpose well with those of night 2) which corresponds to $\sim$82.7 or 84.5 cycles for night 1 and night 2 period, respectively.  Taking into account all the uncertainties, $P$=(12.995$\pm$0.002)/N with 82$\leq$N$\leq$85 or 0.152859 $\leq P \leq$ 0.158500 days i.e. 3.67$\leq P \leq$3.80 hr. This agrees with the periods and uncertainties in Table~\ref{rvsfit}. For the reminder of the paper we adopt as the system orbital period that corresponding to exactly 83 cycles: $P$=12.995/83 days$\sim$3.76 hr. 

\begin{table}
\caption{S-wave radial velocity fits. The $\phi_{R/B}$ is computed with respect to the mid MJD of the first UVB exposure of night 1, i.e. MJD$_1$=58587.011167334.} % (UVB arm, 58587.011227524 for the VIS arm).}
\label{rvsfit}      
\scriptsize
\centering          
\begin{tabular}{lcccc}     % 7 columns 
\hline\hline       
lines & $\gamma$ & K/sin~i & $\phi_{R/B}$ & P \\
 & (km/s) & (km/s) & (MJD-MJD$_1$) & (d) \\ \hline                
 & \multicolumn{4}{c}{night 1} \\ \hline
all & \bf -43$\pm$2 & \bf -178$\pm$3 & \bf 0.1227$\pm$0.0005 & \bf 0.1571$\pm$0.0009 \\
H~I & -55$\pm$4 & -170$\pm$5 & 0.1241$\pm$0.0010 & 0.1562$\pm$0.0015 \\
He~I & -38$\pm$3 & -183$\pm$3 & 0.1222$\pm$0.0005 & 0.1575$\pm$0.0010 \\
C II & -41$\pm$4 & -157$\pm$6 & 0.1232$\pm$0.0015 & 0.1565$\pm$0.0016 \\
\hline                  
 & \multicolumn{4}{c}{night 2} \\ \hline
all & \bf -44$\pm$2 & \bf 170$\pm$3 & \bf 13.1949$\pm$0.0005 & \bf 0.1537$\pm$0.0008 \\
H I  & -67$\pm$5 & -167$\pm$4 & 13.1181$\pm$0.0012 & -0.1562$\pm$0.0019 \\
He I & -32$\pm$2 & -182$\pm$3 & 13.1179$\pm$0.0007 & 0.1563$\pm$0.0013 \\
C II & -45$\pm$3 & -158$\pm$3 & 13.1162$\pm$0.0010 & 0.1582$\pm$0.0014 \\
\hline
%$\dagger$ 
\end{tabular}
\end{table}

We ascribe the narrow s-wave emission to regions near the L1 point on the secondary star in view of its regular pattern, its reduced visibility when it moves from maximum blueshift to maximum redshift, and its velocity width. 

Looking at Table~\ref{rvsfit}, while it seems reasonable and statistically correct to combine the various transitions from different species to produce a single best fit per night, we note that the best fit for each ion separately produces larger $K$ values for He I than the H lines, which have again a greater $K$ than C II 4267.  This arises from two observational biases.  First, not all s-waves are complete and are generally weaker in the branch from maximum blueshift to maximum redshift.  Second, the s-wave width is not the same for all ions, especially at the velocity extremes.  This suggests that the bulk of emission of a given transition is sited differently around the L$_1$ point, perhaps occupying different fractional areas of the sub-stellar hemisphere.  Similarly, the systemic $\gamma$ velocity depends on the ion, $\gamma$(H I) $> \gamma$(C II) $> \gamma$(He I), and  is possibly the result of a velocity gradient in the secondary star "chromosphere". 

The identification of mean location of the s-wave with the L1 point, together with its radial velocity curve and the period, provides a mean for  deriving the binary system parameters: mass ratio, secondary star mass (for an assumed WD mass) and separation. From the relation valid for the center of mass, $a_1M_1=a_2M_2$, and the definition $a=a_1+a_2$, where $M_1$ and $M_2$ are the primary/WD and secondary/donor star mass, while $a_1$ and $a_2$ are the distance of each star from the binary center of mass, and $a$ is their separation: 
\begin{equation}
K_2= \frac{2\pi}{P}\left(a_2-R_2\right) = \frac{2\pi a}{P}\left((1+q)^{-1}-\frac{R_{RL}(2)}{a}\right)
\end{equation}
with $P$ being the orbital period, and $R_{RL}(2)$ the Roche lobe radius of the secondary star for which we use the Eggleton (1983) approximation: 
\begin{equation}
R_{RL}(2)/a=\frac{0.49q^{2/3}}{0.6q^{2/3}+ln(1+q^{1/3})}    
\end{equation}  
If the orbital inclination of the binary is the same as that of the ejecta, $i \approx 40 \deg$, $K_2\approx 270$ km s$^{-1}$ and for a white dwarf mass $M_1=1.2 M_\odot$ (Jos\'e et al. 2020) we obtain $q\gtrsim 0.17$ and a secondary mass $M_2\approx0.21 M_\odot$. 
Alternatively, we can use the orbital period and the $P$-$M_2$ and $P$-$R_2$ empirical relations in Knigge et al. (2011) to constrain the secondary star mass and radius. We find $R_2\approx 0.36 R_\odot$ and $M_2 \approx 0.29 M_\odot$. Consistent results ($R_2\approx 0.39 R_\odot$ and $M_2 \approx 0.34 M_\odot$) are found adopting the Howell et al. (2001) theoretical relations. Hence, for $M_1=M_{WD}=1.2 M_\odot$, $0.24 \leq q \leq 0.28$ and $a\approx 1.4 R_\odot$. 
For the derived parameters the secondary star roughly matches an  M3.5V-M4.5V spectral type (Torres et al. 2010). These are the spectral type we employ in the analysis in Sec.\ref{components}.

\subsubsection{The fuzzballs and the multi-emission simulation}\label{fuzz}

\begin{table}
\caption{Fuzzball-wave radial velocity fits. The $\phi_{R/B}$ is computed with respect to the mid MJD of the first VIS exposure of night 1, i.e. MJD$_1$=58587.011227524. We measured two lines (He~I$\lambda$6678 and He~I$\lambda$7065) on night 1, and only 1 (He~I$\lambda$6678) on night 2; hence the larger uncertainties in night 2 best fit. The orbital period was fixed to $P=0.156566$ days from the s-wave radial velocity fits.} % (UVB arm, 58587.011227524 for the VIS arm).}
\label{fuzzfit}      
\scriptsize
\centering          
\begin{tabular}{lccc}     % 7 columns 
\hline\hline       
lines & $\gamma$ & K/sin~i & $\phi_{R/B}$ \\
 & (km/s) & (km/s) & (MJD-MJD$_1$) \\ \hline                
 & \multicolumn{3}{c}{night 1} \\ \hline
He~I & -55$\pm$4 & -308$\pm$5 & 0.0322$\pm$0.0008  \\ \hline                  
 & \multicolumn{3}{c}{night 2} \\ \hline
He I & -74$\pm$10 & -288$\pm$9 & 13.1898$\pm$0.0015  \\ \hline
\end{tabular}
\end{table}
The trailed spectrograms of the VIS arm He~I transitions suggest that the fuzzballs are not a confined emission region that appears only at unique orbital phases, but is always visible during its orbital motion around the binary center of mass. This is even more evident in the trailed spectrograms of the H Paschen and Brackett series (see Fig.\ref{trails}). The superposition of the emission from the fuzzball-wave with the that from the ejecta simply makes it more visible at maximum blueshift and redshift in the bluer H-Balmer transitions. 
Measuring the fuzzball-wave of the VIS arm He~I lines\footnote{We did not measure the radial velocity of the H Paschen and Brackett transitions because of their width and low S/N (especially when telluric  absorptions or poorly removed sky emission lines are superposed).} and fitting their radial velocity using bootstrapping techniques as we did for the L1 s-wave, we obtain the radial velocity parameters in Table~\ref{fuzzfit} and Fig.\ref{rv}. The fuzzball-wave is not exactly in antiphase with the L1 s-wave. So although, it arises from a region within the WD Roche Lobe it is not symmetrically distributed around the WD and cannot trace the WD orbital motion. The exact phase  difference is $\phi_{RB}(s$-$wave)-\phi_{RB}(fuzzball$-$wave)\simeq$~0.58 periods, %%0.578095$
meaning that the fuzzball-wave extremes (i.e. max blue/red shifts) anticipate those of the L1 s-wave by $\sim0.08$ orbital periods.
Within the standard convention where phase 0 is the secondary inferior conjunction\footnote{Note that $t_0$ defined in the previous section correspond to phase 0.5 in this convention.}, the secondary star L1 emission reaches maximum blue/red shift at phase 0.75/0.25 as expected. The fuzzball region, instead, reaches maximum blue/red shift at phases  0.18/0.68. Red to blue crossing occurs at phase $\sim$0.93. The phase shift suggests that the fuzzball emitting region could arise from the stream impact point and that we see maximum red-/blue-shift when we look up-/down-stream. 
%

%----------------------------------------------------------------- 
   \begin{figure} 
   \centering
\includegraphics[width=6.5cm]{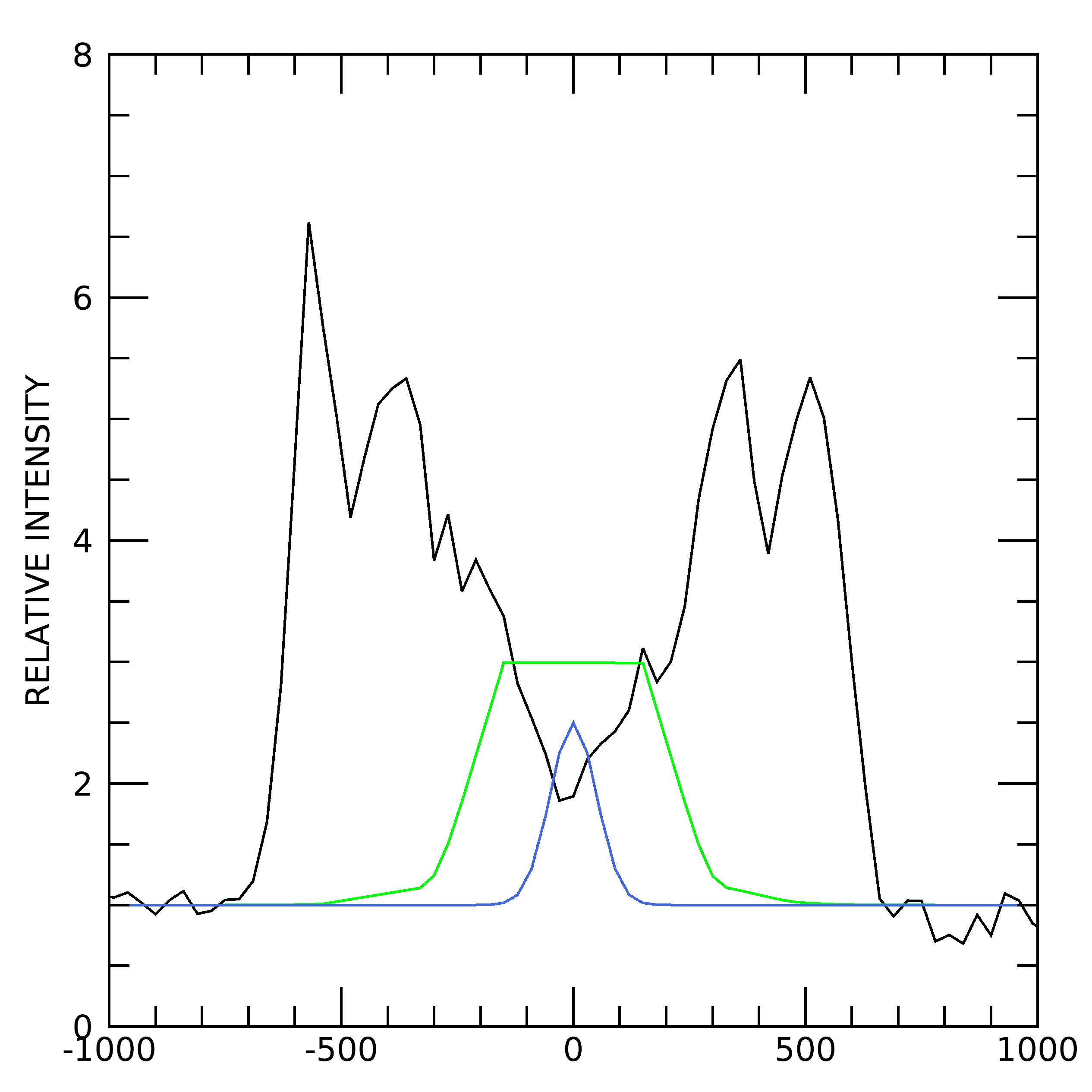}
\includegraphics[width=6cm]{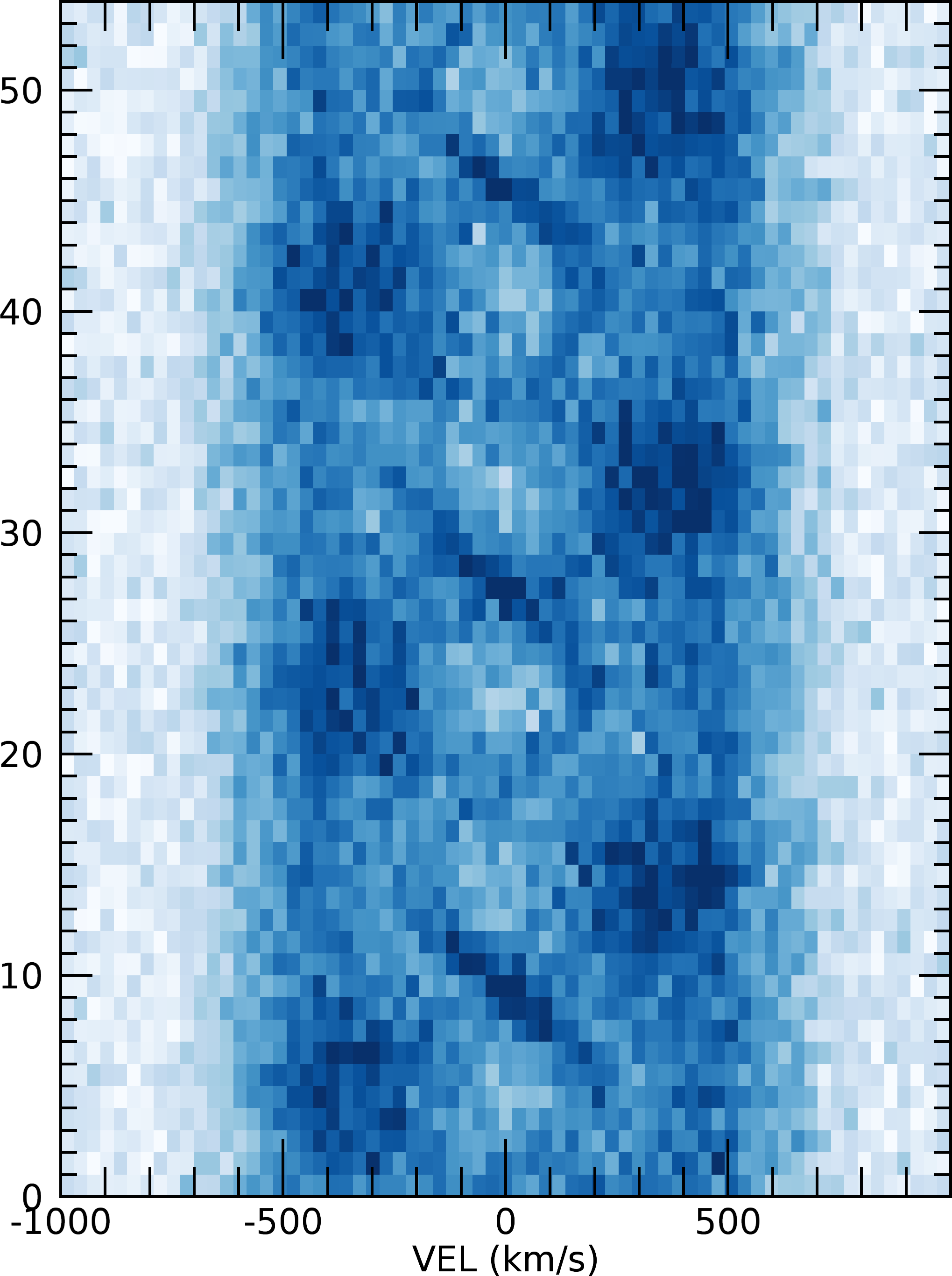}
   \caption{Top: Cen nebular line ([Ar~III]$\lambda$7135, black line), together with the Gaussian (blue) and the "boxy" (green) profiles used in our simulation. Bottom: The simulated trailed spectrum. See Section~\ref{fuzz} for details.
              }
         \label{simu}
   \end{figure}
%-----------------------------------------------------------------
Fig.\ref{simu} shows a simulation produced by superposing a nebular line profile with a narrow ($\sim$50 km/s) Gaussian emission and a broader ($\sim$300 km/s) boxy emission with the phase offset mentioned above and with a velocity amplitude matching the best fit $K$ of the s- and fuzzball-wave, respectively. The simulation qualitatively reproduces our observations, supporting the contention that the fuzzballs arise from an emitting region within the WD Roche Lobe that is continuously visible around the orbit. We do not obtain significant differences replacing the boxy profile with a different one (e.g. a disk/double peaked profile). Thus, although we cannot definitively constrain the geometry and ultimate nature of the emitting region, the fuzzballs source cannot match an accretion disk because of its asymmetric location, phasing, and $K$ value. Instead, it can match a stream impact region within the circum-WD matter. 

Since we are not aware of any cataclysmic variable (CV) with an optically thin hot-spot (intended as the classic stream-disk impact point) and an optically thick disk (e.g. Warner 1995, see also Sect.\ref{discussion}) we further checked for possible accretion disk signatures in the emission lines wings. Specifically, we inspected the trailed spectrogram after enhancing the contrast at the wings level and searched for any coherent motion using the bisector technique for the wings of isolated and relatively strong emission such as H$\beta$, H$\varepsilon$, and He~II (and [OIII] for consistency). We found values randomly scattered within $\lesssim$40 km/s. The expected WD orbital velocity amplitude for the system parameters in  Sect.\ref{rvmass} is $\sim$60 km/s, i.e. the probed quantity would be at the limit of our precision. 

We also note that the emission lines have FWZI$\sim$1300 km/s implying that they cannot probe any region close to a WD at its expected equilibrium radius such as the inner accretion disk. This also excludes high velocity wings of the isolated single peaked emission lines that are observed in some SW Sex type stars (e.g. Rodriguez-Gil et al. 2007). 

%-------------------------------------- Two column figure (place early!)
   \begin{figure*}[h]
   \centering
   \includegraphics[width=19cm]{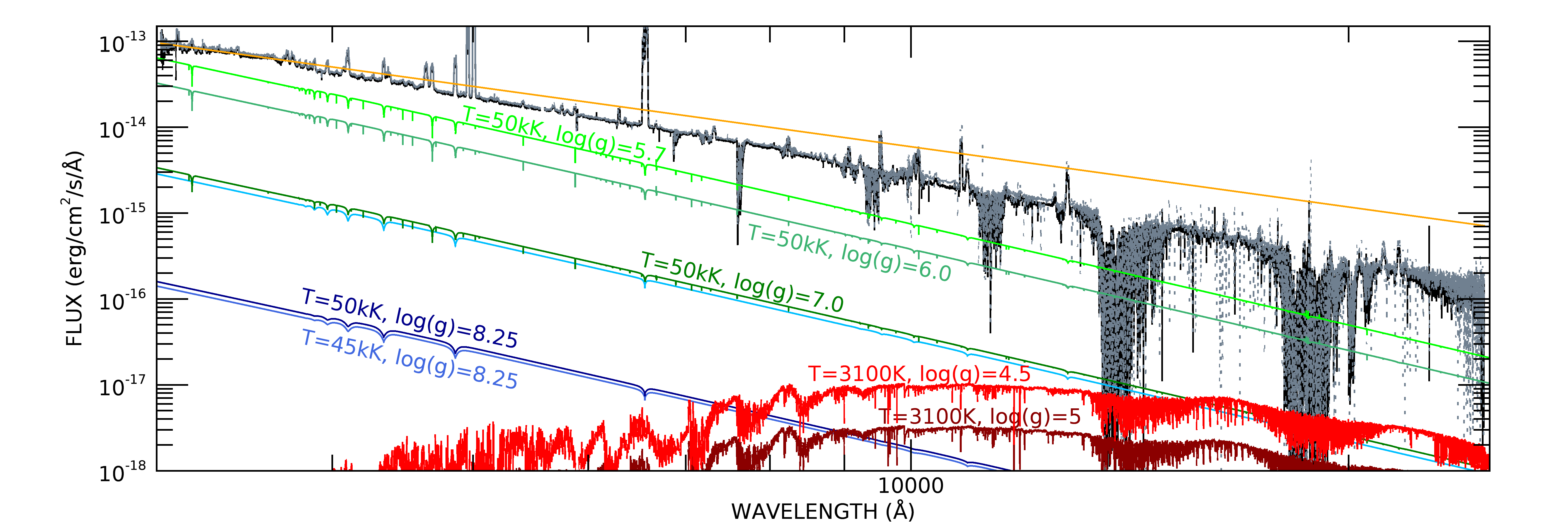}
   \includegraphics[width=19cm]{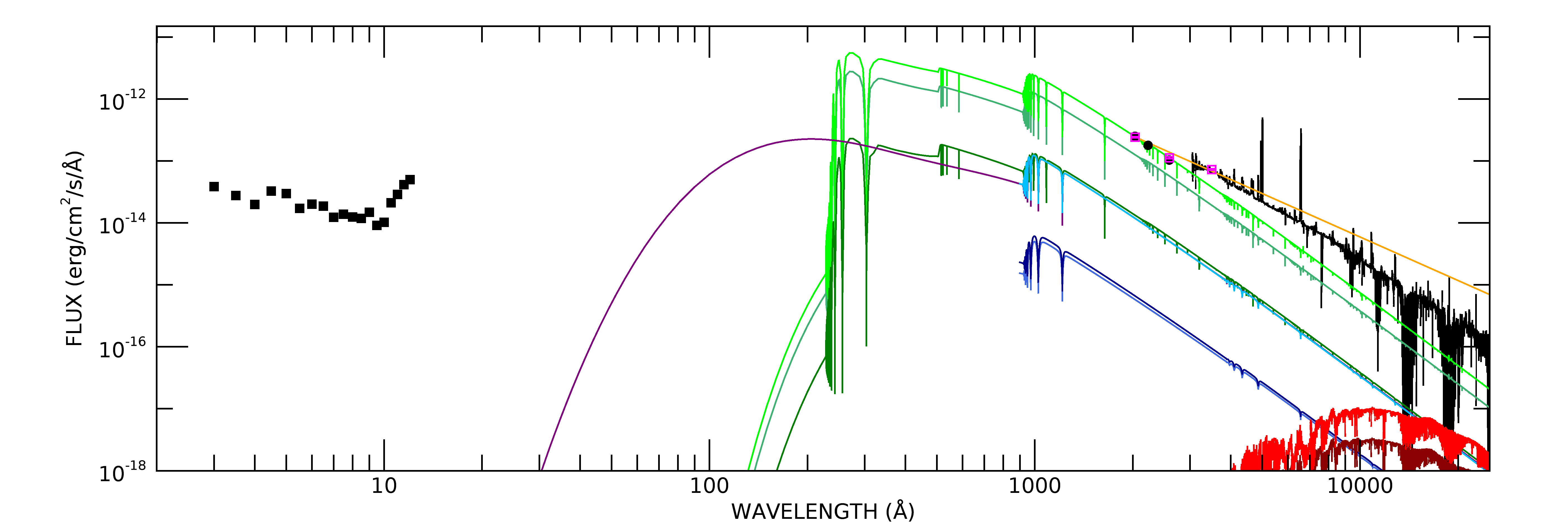}
   \caption{Cen continuum SED and comparison with theoretical models. Top panel: The black solid line is for the Cen fiducial reference spectrum 4 from night 1; while the grey dotted one is for the fiducial spectrum 8 from night 2. Both have been corrected for reddening. The orange solid line represents a power-law of power -7/3 as predicted for steady state accretion disks; while all the other colored lines are stellar atmosphere models. Their temperature and gravity are marked in the figure itself (shade of blues for the Koester WD models; shades of green for the Tubingen database stellar atmospheres; shade of red for the Next Gen NLTE cool stars models). Bottom panel: The fiducial reference spectrum 4 from night 1 together with the {\em Swift} UVOT photometry (black filled circle for the observation of April 2019; magenta open squares for the observations of June 2019), the {\em Chandra} spectrum, the same models than in the top panel, and (magenta line) a T=50kK $\log g$=7.0 model with solar abundances (Tubingen database). 
   }
              \label{seds}%
    \end{figure*}
\subsection{The continuum spectrum}\label{cont}
Since for most of the night the Cen observations were obtained at a better seeing than the spectrophotometric standard star EG~274, the Cen spectra tend to be overestimated in the blue flux. Selecting the spectra that have UVB and VIS PSF roughly matching that of EG~274 we can be confident that their SED is accurate. However, the Cen spectra having PSF matching or comparable to that of the spectrophotometric standard in the UVB and VIS arm might have non-matching PSF in the NIR arm. This is because in addition to the optics and the seeing,   
the NIR PSF also depends on sky conditions such as the humidity and the number and size of aerosol. 
Thus, we chose spectrum number 4 for night 1 and  spectra numbers 8 and 42 for night 2 as fiducial reference spectra for inferences about the Cen SED (Sec.\ref{components}) and its long term evolution (Sec. \ref{long-tv}) since they were obtained at seeing comparable to that of the spectrophotometric standard. The spectra, although from different nights and at slightly different orbital phase, are consistent with the same SED. Their ratio is constant and they differ in flux by $\sim$6\%, 5\% and 13\%, in the UVB, VIS and NIR arm, respectively. 

\subsubsection{The SED and its components}\label{components}

Figure~\ref{seds} shows the Cen SED after correction for interstellar reddening, E(B-V)=0.15 mag (Mason et al. 2018). The top panel, zooms in the two XShooter fiducial spectra, while the bottom panel shows the complete data-set from the x-ray to the NIR wavelength. 
We compared the data with a number of stellar photosphere models and an arbitrarily scaled power-law expected for an optically thick accretion disk in steady state (e.g. Pringle 1981). We selected stellar models that most closely matched the two stars composing the binary system. For the donor star we chose a 3100~K, $\log g$=5 or 4.5, solar abundance model (red lines in the figure) from the Next Generation database (Allard et al. 1997, Hauschildt et al. 1999 and references therein), consistently with our results in Section~\ref{rvmass} and Torres et al. (2010) calibrations. 
For the primary we selected models from the Koester WD database (Koester 2010, Trembley \& Bergeron 2010) and the Tubingen NLTE model atmosphere database (Rauch et al. 2018). The Tubingen models are He rich (50\% by number). The temperature was limited to the range 40000-50000~K based on the observed s-wave transitions, assuming that the L1 region is excited by the WD. Since we detect only H, He~I, and C~II transitions and not He~II or Ca~II, the bulk of the source radiation must be confined within 10 and 50 eV, i.e. consistent with a T$_{eff}\leq$50000 K photosphere. $\log g$ was set to 8.25 as appropriate for a $\sim$1.2 M$_\odot$ WD (shade of blue lines).
Both the donor and the WD models have been scaled to the distance of 2 kpc. 

It is immediately clear why we detect no signatures from the secondary star: its fractional contribution is at most a few percent of the total observed flux. For the same reasons, a normal WD cannot be detected, even at UV wavelengths. Both the WD and the donor are outshone by something else. 
It is also evident that the SED slope does not quite match the steady state accretion disk slope. It is not rare that a CV SED deviates from that of an optically thick accretion disk (see  Sect.\ref{discussion}), although best matches are found, and expected, for disks with high mass transfer rates (e.g. SS~Cyg during outburst, Kiplinger 1979, see also Warner 1995).  
Linear fits to the Swift photometry and points on the opt+NIR continuum show that the Cen SED is better represented by a broken power law with power $-$2.4 in the NUV range (in good agreement with the optically thick steady state accretion disk), and a steeper slope (power $-$3.2) longward of $\sim$4000\AA. 

We can proceed making two different alternative hypothesis: 1) the dominant emitting source is the accretion disk, both in the optical and the NUV; 2) a bloated WD photosphere is the major continuum source in this wavelength range. 

For hypotheses 1) we can compute the NUV+opt+NIR luminosity and use it to constrain the accretion disk mass transfer rate. The dereddened flux of the least overlapping {\em Swift} UVOT filters is F$_{uvw2+uvw1}$=2.4$\times10^{-10}\pm1.1\times10^{-13}$ erg/cm$^2$/s, and the average integrated flux of the dereddened fiducial spectra after removal of the emission lines, F$_{UVB+VIS+NIR}$=2.2$\times$10$^{-10}\pm$4.8$\times$10$^{-12}$ erg/cm$^2$/s. This gives a luminosity L$_{NUV - NIR}$=1.8$\times10^{35}$ erg/s $\simeq$47 L$_\odot$. Then, approximating \.M$\sim$2LR$_{WD}$/(GM$_{WD}$) with the usual meaning of the symbols, and adopting M$_{WD}$=1.2~M$_\odot$ (hence R$_{WD}\simeq 0.0115$ R$_\odot$), we get the mass accretion lower limit log(M$_\odot/yr$)$\sim-7.5$  which is  larger than what is typical for CV above the period gap (Howell et al. 2001, Knigge et al. 2011) or measured in old nova systems\footnote{A direct comparison, however, is not possible since Selvelli and Gilmozzi computed the \.M over a different wavelength range taking into account inclination and limb darkening effects.} (Selvelli and Gilmozzi 2019). 
It must be noted that this is a conservative lower limit since we do not have a measure of the FUV flux where most of the accretion disk luminosity is expected to emerge. 
Using the above power-law fit to the Swift photometry and extending it to the IUE range (1250-3050~\AA) we can estimate the Cen UV luminosity and compare it to  that (uncorrected for geometrical effects) of the old-novae in Selvelli and Gilmozzi (2019). We find L$_{IUE}\sim$63 L$_\odot$ which is larger than the values in Selvelli and Gilmozzi's Table~2, as we will discuss in Sect.\ref{discussion}.

For hypothesis 2) we can check the lowest $\log g$ that is compatible with the data. Figure~\ref{seds} shows that He rich photosphere model of T=50000~K and $\log g$=5.7 (lime line) can match our data and the Swift NUV photometry as well as the power law approximation of the optically thick accretion disk. Models of intermediate $\log g$ between this upper limit and that appropriate for an equilibrium WD are also plotted for comparison. 
The bloated WD photosphere with $\log g$=5.7 would be as large as $\leq$0.25 R$_\odot$, i.e. almost $\sim$20 times a normal WD radius and $\sim$0.3 R$_{RL}$(WD). As unconventional as this may appear, it is an alternative to the almost Roche lobe filling accretion disks proposed for similarly bright systems (e.g. HR Del, Bruch 1982). In addition, the NUV slope, so similar to that of a stellar photosphere model, together with the flatter opt+NIR continuum is consistent with an expanded WD envelope whose cooler outer layers are more opaque than the inner ones and responsible for a flatter optical slope. Any possible absorption lines from such an expanded WD envelope, if present, are masked by the emission lines from the rest of the binary and the ejecta. 
\subsubsection*{The WD accretion rate from the x-ray data}
The {\em Chandra} spatial resolution is poorer than the estimated nebula size in the optical so it does not resolve the ejecta. Nevertheless, the x-ray signal detected with {\em Chandra} certainly originates in the binary and not from the ejecta. Only a few novae, several decades to century old, have shown x-ray emission due to interaction between the interstellar medium (ISM) and the ejecta (e.g. Takei et al. 2013 and reference therein). Cen is not located in a particularly dense interstellar region and it is still too soon after its eruption to have accumulated enough ambient material to be detectable in the x-rays.

Cen has an integrated flux of F$_X\simeq 2.2\times$10$^{-13}$ erg/s in the range 7-12.4 \AA \ (about 1.0-1.8 keV), corresponding to the luminosity L$_X$=1$\times$10$^{32}$ erg/s for the adopted distance of 2kpc. Although a direct comparison with published x-ray and optical observations of other bright CV is not straightforward because of the slightly different energy and wavelength ranges, we note that the derived luminosity is in line with that derived for other CV and nova-like systems (e.g. Balman 2020 and reference therein) and the ratio between the x-ray and the optical flux is consistent with that observed in high mass transfer rate systems (e.g. Kuulkers et al. 2006 and references therein). 

Where the x-ray originate in the system cannot be definitively assigned at this stage.  %\footnote{The integral flux and the total luminosity are F$_X$=2.6$\times{-13}$ erg/cm$^2$/s and L$_X$=1.24$\times$10$^{32}$ erg/s fitting a model in the range 0.5-7 keV}. 
Considering the error bars and the low source  counts, the x-ray signal is consistent with no modulation over the orbital period, indicating that the emitting region is always in view. 
It could be from the WD boundary layer, which yields an estimated mass accretion rate of log(M$_\odot/yr)_X\sim-10.8$,  that is almost 3 order of magnitude smaller than that derived from the optical SED. Alternatively, it could arise from any of the additional regions hypothesized to explain x-ray/opt flux dichotomy (Kuulkers et al. 2006 and reference therein, see also Mukai 2017 and Balman 2020).  
However,  such a low x-ray emission could originate at an impact site within the WD Roche lobe, with the accretion onto the WD being extremely inefficient and therefore remaining undetected. 

\subsubsection{Short term SED variations: intra and across nights}\label{short-tv}

In the attempt to construct a continuum light curve from the spectra, we first tried to mitigate the color effects induced by the seeing and the flux calibration (see beginning of Section~\ref{cont}). We determined the wavelength dependence of the spatial PSF (each XShooter arm is a separate cross dispersed echelle spectrogram) by measuring the width of the spectrum at three points along each separate order. We then fit a quadratic function after normalization to a selected wavelength in each arm, and used the fit to roughly correct for slit-losses.
Such a correction is appropriate for relative comparisons of the continuum flux within an arm. It is not a method for absolute flux calibration and SED reconstruction.  

The differential photometry was extracted from portions (10 to 60 \AA\ wide) of the continuum and then compared, for consistency, with the integrated flux of nebular lines, keeping in mind that the continuum source and the nebular lines are not necessarily co-spatial (see Sect.\ref{sec:2d}). We found similar light curve patterns in the continuum and the nebular lines, and  thus conclude that the source flux is constant to within the precision of our method ($\sim$15, 25 and 20 percent in the UVB, VIS and NIR arm, respectively). 
The lack of any intrinsic variability in the K band is particularly interesting since it supports our results from Sec.\ref{components}, indicating that tidal distortion of the secondary star is outshone by a stronger light source, and that  high energy radiation incident on the WD-facing side of the donor does not penetrate deep into the stellar photosphere and affects only its upper layer. 

\subsubsection{SED long term evolution: from late nebular to now}\label{long-tv}
   \begin{figure*}
   \centering
   \includegraphics[width=18cm]{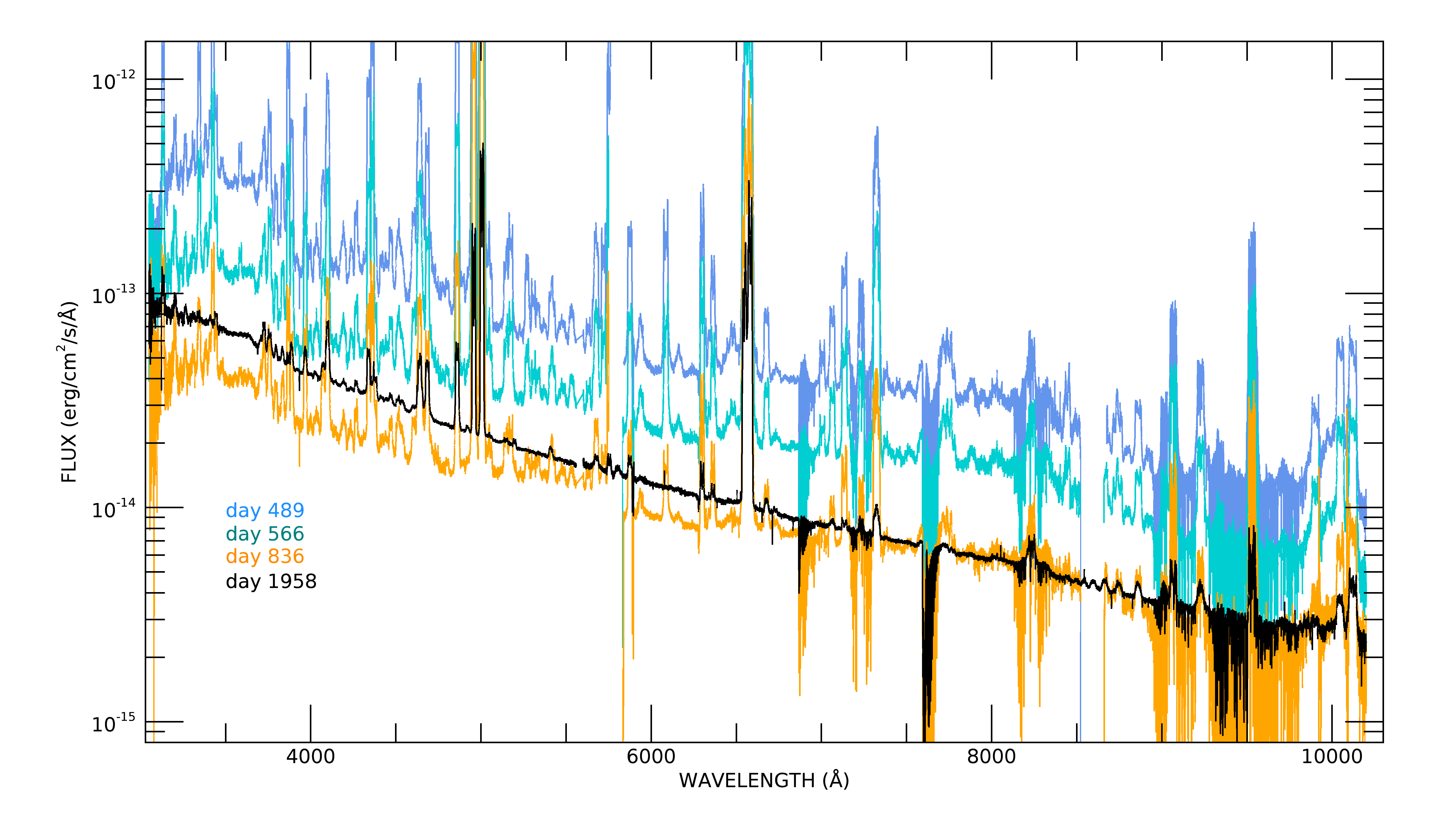}
      \caption{Comparison of the UVES spectra with our fiducial spectrum from night 1. Note that since UVES spectra do not extend into the NIR wavelengths, XShooter NIR is not shown. The spectra have been corrected for interstellar reddening.
              }
         \label{longTermSED}
   \end{figure*}

The comparison of our fiducial reference spectra with those from previous epochs provides information about the evolution of the SED. Figure~\ref{longTermSED} displays the optical part of the fiducial reference spectrum 4 on night 1 together with the UVES spectra taken when the nova was at $\sim$1.3, 1.6, and 2.3 years after outburst.

\begin{figure}
   \centering
   \includegraphics[width=9cm]{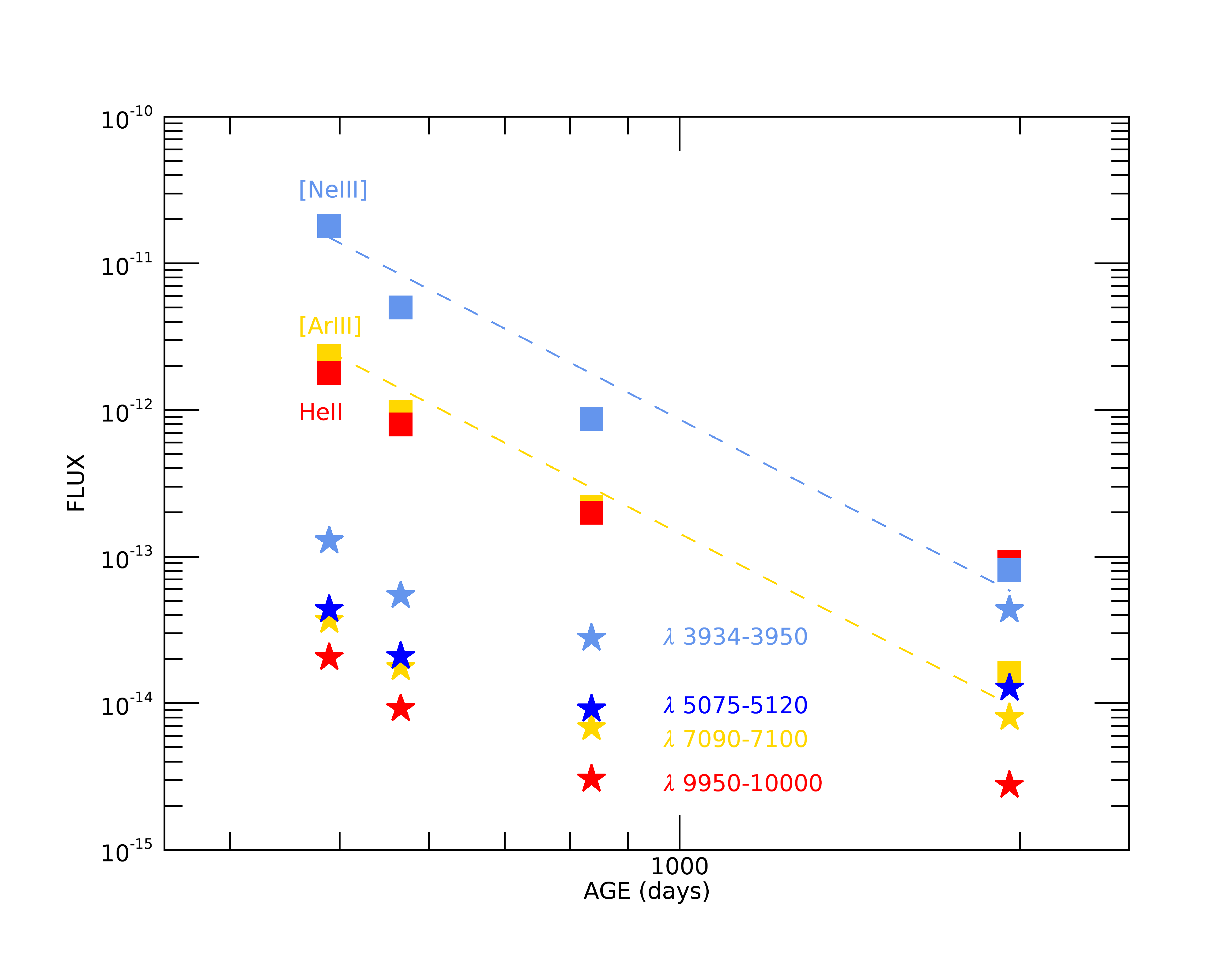}
      \caption{Flux decline rate from day 489 to 1958 as measured in portion of the continuum (star symbols) and in selected emission lines (square symbols), two nebular ([Ne III]$\lambda$3869 and [Ar III]$\lambda$7135) and a permitted one (He~II $\lambda$10124). 
              }
         \label{lt_flux}
   \end{figure}

The absolute flux calibration of the UVES spectra is not sufficiently accurate and the blue part of the spectra is likely underestimated, since the flux calibration was performed with the observatory standard star which is observed with a much wider slit than the Cen.
We verified that using passbands described in Bessel (1976), the derived synthetic photometry was usually consistent with the AAVSO (prevalidated) V band photometry but significantly fainter than the B band photometry at the same epochs. Differences as large as 1.4-1.5 magnitudes point to systematics that can be significant, especially for the B band (Munari and Moretti 2012).  In addition, the AAVSO photometry can vary among observers and were typically from only one observer per epoch. Consequently, we did not adjust the UVES data to the AAVSO photometry.  
Nevertheless, the UVES spectra are similar to each other and different from that with XShooter in the more pronounced continuum excess below $\sim$4000~\AA\ and possibly below $\sim$8500 \AA. This suggests stronger continuum radiation from recombination (mainly H-I free-bound transitions) from the ejecta at the time of the UVES observations. However, the XShooter spectrum suggests that some recombination continuum also comes from the binary system and the ongoing accretion. To better understand the relative contribution of the ejecta and the binary, we measured the continuum flux at different portions of the spectrum along with the integrated flux of a few emission lines. 
Fig.\ref{lt_flux} shows how the continuum and emission lines fluxes decrease. It is evident that the forbidden transitions decline roughly as a power law in time, consistent with forming in an environment of decreasing density and in the absence of any new photoionizing source (Shore et al. 2016). In contrast, the continuum flux and the He~II transition flatten out indicating that a new source is contributing (in addition to recombination from the ejecta). 

\section{Discussion and Conclusions}\label{discussion}

Cen is a typical classical nova. It was only because of its brightness that it was so extensively observed from maximum to late decline up to 5.4 years after the outburst. What makes this nova special is the collected data set that is not available for other systems: multi-epoch UV and optical high resolution spectroscopy from early decline to late nebular stages, {\em Swift} x-ray monitoring during the whole outburst, and, now, medium resolution time resolved optical+NIR spectroscopy complemented with {\em Swift} and {\em Chandra} observations. 
The collected observational evidences should, therefore, be representative of any classical nova of intermediate speed class and inclination. 

Observationally, we identify two sinusoidal patterns that vary in anti-phase. One is a narrow sine-wave with velocity amplitude $\sim$178 km/s (the s-wave). It is fairly narrow, and has varying visibility, being more evident during the red-to-blue branch of the radial velocity curve. This phenomenology suggests that it can be assigned to the irradiated face of the secondary star. In particular, since the line width is narrow, the emitting region is fairly small --that is, not extended over the whole hemisphere facing the WD-- and is probably confined to the proximity of L1. We exclude that the narrow s-wave is from the stream (other than possibly the very initial part near L1) because the emission line width would be greater (due to the velocity gradient), and it would have a different visibility pattern. More important, the stream has too low a density --and therefore a negligible emission measure-- to be the dominating source. 

Irradiation of the secondary star is observed in cataclysmic variables (e.g. Catalan et al. 1999, Harrison et al. 2003, Southworth et al. 2008, Schmidtobreick et al. 2018) and it is expected in novae for which increased mass transfer rate is predicted as consequence of the expected high WD temperature (Shara et al. 1986, Kovetz et al. 1988). We note, however, that the observed irradiation is not quite as modeled by Barmann et al. (2004; and specifically for the cataclysmic variable VV~Pup in Mason et al. 2008). Cen displays  only emission lines, while the continuum remains unaffected by the incident radiation; the incident high energy flux on the donor is not penetrating deep into the stellar photosphere and only produces a chromosphere-like emission in  its upper layer. The temperature inversion caused by the irradiation is, however, different from a typical chromosphere since we detect transitions from higher ionization potential energy ions than in typical stellar chromospheres, namely H, He~I, and C~II without Na~I~D and Ca~II. The detected species constrain the ionizing source energy distribution. In particular, the lack of singly or doubly ionized metal lines and He~II suggests an irradiating source close to $\sim$50000 K in color temperature.  We propose that the irradiating source is the WD which is, however, much more extended than expected from the equilibrium mass-radius relation.
From this s-wave we derive the system orbital period and, with few assumptions, system parameters such as the secondary star mass and radius, the mass ratio (see Section 3.2.1).

The second sinusoidal pattern is the one we nick-named "fuzzballs" because of its very localized visibility in the upper H Balmer series lines. However, its periodic visibility is simply due to the superposition of different contributions: the fuzzball source and the ejecta. As already mentioned, the fuzzballs morph into a complete sinusoidal pattern in the Paschen and Brackett series. 
The amplitude and phase of the fuzzball-wave well match the position of an impact region. The region appears to be optically thin gas because its emission is not modulated with orbital phase. 

In the conventionally adopted picture of CV (Warner 1995 and references therein), the systems with low mass transfer rate host an optically thin disk (at least partially) that generates emission lines together with stream-disk impact region (the hot-spot). Instead, high mass transfer rate systems tend to have optically thick disks with weak or no emission lines and absorptions. These systems also do not show emission lines from the hot-spot, since it emits mostly in the continuum like the disk. 
If we interpret the fuzzball-wave as the hot-spot, we face an anomaly since we see emission lines from the impact region but not from the disk. However, SW Sex are similarly anomalous systems with just a single peak emission producing a relatively large velocity amplitude sinusoidal wave (order of few hundred km/s), and occasional evidence for an irradiated face of the secondary star. The identifying characteristics of the class (Rodriguez-Gil et al. 2007) are maximum blueshift of the sinusoidal wave at orbital phase 0.5, transient absorption features at about the same orbital phase, high velocity wings (even up to few thousands km/s) whose radial velocity curve is offset in phase (by 0.1-0.2) with respect to that of bulk of the emission, and, as noted by Dhillon et al. (2013) flux variation of the emission lines driving the broad band photometry variability. These conditions need not be simultaneously fulfilled to make an SW Sex star (even the prototype itself happens to lack many, Dhillon et al. 2013 and reference therein), but Cen has none of these. 
The anomaly extends to the continuum emission. Although multi-band studies of bright CV SED are not numerous, Kiplinger (1979) has shown that the SED of the SS Cyg accretion disk during outburst matches the theoretically predicted spectral energy distribution of an optically thick geometrically thin accretion disk from the FUV to the mid-IR. Similarly, Ciardi et al. (1998) found that the accretion disk of the nova-like (NL) V592 Cas matches a power-law distribution with power $-$2.3 from the U to the NIR, in excellent agreement with the Pringle (1981) approximation. In contrast, Hoard et al. (1997) report a flat optical continuum (in wavelength) for the SW Sex stars. 

The SED analyses of the accretion disk in bright CV have been generally limited to the UV wavelength because this is where the most of the accretion disk flux is emitted. Puebla et al. (2007), analyzing UV spectra (IUE and HST) of NL and old-novae (i.e. systems supposedly hosting a stable accretion disk), found that that optically thick geometrically thin Keplerian disks models cannot fully reproduce the observations and suggest a number of scenarios that might be explored to improve the fitting and remove degeneracy. Selvelli and Gilmozzi (2013) used IUE, FUSE and HST/STIS spectra of old-novae, fitting them with a power-law. They found a different power for each nova, on the whole in the range $-$0.32 to $-$2.55 or $-$0.35 to $-$2.88, depending on the adopted reddening law. Gilmozzi and Selvelli (2021) found a similar range of powers ($-$0.15 to $-$2.36) analyzing the IUE spectra of NL (including SW Sex systems). Selvelli and Gilmozzi (2019) and Gilmozzi and Selvelli (2021) computed also the IUE luminosity for each object using the Gaia releases. Even if the Cen UV slope is consistent with that of an optically thick accretion disk and in the range observed for old novae and NL the difference between Cen and NL or old novae becomes more obvious when the fluxes are converted to luminosity. The average luminosity of NL systems is $\sim$3$\pm$2 L$_\odot$, that of old novae is larger and more scattered $\sim$6$\pm$7 L$_\odot$ depending on the age of the old novae (17-29 to 132-147 yr after outburst). Both results exclude the anomalously bright objects, QU Car (for the NL) and HR Del (for the old novae), whose luminosities (40 and 64 L$_\odot$, respectively) are much closer to the 63 L$_\odot$ estimated for Cen in Sect.\ref{components}. 
The reason to treat those objects separately are the debated nature of QU Car and the relatively young age of HR Del at the time of the IUE observations (12-25 yr). QU Car has been suggested to belong to the V Sge stars class (Oliveira et al. 2014) and to be one of those permanent super soft source binaries that miss detection in the X ray because of high interstellar extinction. HR Del was an extremely slow nova with an anomalous light curve that could result from unstable burning, as proposed for the recent transient ASASSN-17hx (Mason et al. 2020), and with a similarly  extremely large ejecta mass (9$\times$10$^{-4}$M$_\odot$, Moraes and Diaz 2009). In other words, both objects are neither typical nor representative of their own class, and both could have ongoing burning and, therefore, a completely different WD structure. 

The classic {\it default} assumption of a fully reformed stationary accretion disk might be not the most appropriate when confronting of a system that only $\sim$5 years ago experienced a nova outburst, that has an extraordinary luminosity and is still over 1 mag brighter than before outburst. Is the assumption of such a disk is realistic?  
The WD has experienced a TNR and it is still unknown whether a disk survives the blast. Similarly uncertain is how the WD relaxes and on what timescales. It is hard to imagine that after the TNR, the primary Roche lobe is empty except for the WD. Any (He) envelope remaining on the WD surface (e.g. Prialnik 1986, Newsham et al. 2014), should relax on thermal time scales once all nuclear burning has ceased\footnote{The accumulated mass before the outburst is in the range 10$^{-5}$-10$^{-4}$ M$_\odot$ but the ejection efficiency is highly uncertain and debated. The other free parameters in the computation of the thermal time scale (WD mass and radius) are similarly uncertain.  An estimate for the WD radius, in particular, would require dynamically simulating the WD structural evolution from the time of the ejection through its relaxation to equilibrium. Such models are currently unavailable. Nevertheless, depending on the assumed parameters, estimates of the thermal time scale vary from years to century. For Cen, assuming that at least $\sim$3\% of the accreted envelope remained on the WD surface (Prialnik 1986) the thermal time scale is $\sim$4-5 years at the observed luminosity and for WD radii as large as 0.2-0.25 R$_\odot$. However, both the WD luminosity and radius are a decreasing function of time.}. But even whether that stage has been reached so soon is theoretically uncertain. Unfortunately, observations cannot probe the system status during the optically thick phase of the outburst (which depend on the wavelength). Additionally, theoretical models and simulations, to date, have focused on different issues such as the TNR nucleosynthesis, the mixing, the ejecta mass and ultimate growth of the WD to the Chandrasekhar limit (e.g. Starrfield et al. 2016, Jos\'e 2016, Yaron et al 2005, Hillman et al. 2016), and, because of computing time limitations, are one-dimensional. Last but not least, smooth particle hydrodynamic simulations of accretion disks use idealized initial conditions, e.g. an empty Roche lobe (Frank et al. 2002 and references therein).  

Instead, adopting an unusual but not unprecedented scenario of a still expanded WD envelope surrounded by gas (from the TNR leftover and from the secondary) we can consistently explain our observations. The expanded WD photosphere scenario was first proposed by Shore et al. (1997) for the ONe nova V1974 Cyg. There, UV spectra taken $\sim$3 yr after outburst showed that the Ly$_\alpha$ absorption from the primary and the continuum were better matched by a $\log g$=6 photosphere than the $\log g$=8 one. While the estimated luminosity was larger than theoretically predicted for a relaxed WD but was consistent with the derived continuum for an effective temperature of 20000 K.

For Cen, the bloated WD envelope can account for most of the system luminosity, the exact fraction depending on the chosen $\log g$. The unsettled gas within the WD Roche lobe, while not excluding density concentrations on the orbital plane, explains the lack of accretion disk signatures since bound material in chaotic motion emits all over the place in any given transition. Finally, the fuzzball-wave, being self-similar at any phase can be explained with the stream impacting optically thin low density gas without developing any gradient and wake in the impact point. 

The proposed scenario is quite different from those for U~Sco, LMC 1968 and nova Her 1991, for which evidence of early accretion disk formation has been reported. While the data sets are not directly comparable, these are either recurrent nova systems or extremely fast CN for which the ejecta masses are orders of magnitude smaller than Cen and the super soft source phase much shorter.

%-----------------------------------------------------------------

\begin{acknowledgements}
EM thanks Yazan Al Momany for unveiling SM secrets. The authors also thank Jordi Jos\'e and Juan Echevarria and for discussion and Pierlugi Selvelli both for discussion and for sharing unpublished results. We are very grateful to the referee, John Thorstensen, for having provided constructive feedback that greatly improved our work. 
\end{acknowledgements}


\begin{thebibliography}{}

\bibitem[allard(1997)]{allard} Allard, F.; Hauschildt, P.H.; Alexander, D.R.; et al., 1997, ARA\&A, 35, 137

\bibitem[bailer(2018)]{jones}Bailer-Jones, C.A.L.; Rybizki, J.; Fouesneau, M.; et al., 2018, A\&A, 516, 58

\bibitem[Balman(2020)]{balman} Balman, S., 2020, AdSpR, 66, 1097

\bibitem[barman(2004)]{barman} Barman, T.S.; Hauschildt, P.H.; Allard, F, 2004, ApJ, 614, 338

\bibitem[bessel(1976)]{bessel} Bessel, M.S., 1976, PASP, 88, 557

\bibitem[Breeveld(2011)]{breeveld}Breeveld, A.A.; Landsman, W.; Holland, S.T.; et al, 2011, AIPC, 1358, 373

\bibitem[bruch(1982)]{bruch} Bruch, A., 1982, PASP, 94, 916

\bibitem[ciardi(1998)]{david} Ciardi, D.R.; Howell, S.B.; Hauschildt, P.H.; et al., 1998, ApJ, 504, 450

\bibitem[cataln(1999)]{catalan} Catal\'an, M.S.; Schwope, A.D.; Smith, R.C., 1999, MNRAS, 310, 123

\bibitem[Dhillon(2013)]{vik} Dhillon, V. S.; Smith, D. A.; Marsh, T. R., 2013, MNRAS, 428, 3559

\bibitem[eggleton(1983)]{eggleton} Eggleton, P.P.; 1983, ApJ, 268, 368

\bibitem[frank(2002)]{frank} Frank, J.; King, A.; Raine, D., 2002, in {\em Accretion Power in Astrophysics}, Cambridge University Press

\bibitem[fuentes(2021)]{fm} Fuentes-Morales, I.; Tappert, C.; Zorotovic, M.; et al., 2021, MNRAS, 501, 6083

   \bibitem[Gaia DR2 (2018)]{gaia} Gaia Collaboration, 2018, A\&A, 616, 1
   
\bibitem[gilmozzi(2019)]{gilmozzi} Gilmozzi, R.; Selvelli, P.L., 2021, to be submitted to A\&A

\bibitem[hanuschik(2000)]{pam} Hanuschik, R.; Amico, P., 2000, Msngr., 99, 6

\bibitem[Harrison(2003)]{harrison} Harrison, T.E.; Howell, S.B.; Huber, M.E.; et al., 2003, ApJ, 125, 2609
   
\bibitem[peter(1999)]{peter} Hauschildt, P.H.; Allard, F.; Baron, E., 1999,  ApJ, 512, 377

\bibitem[hoard(1997)]{hoard} Hoard, D.W.; Szkody, P., 1997, ApJ, 481, 433
   
\bibitem[howell(2001)]{howell} Howell, S.B.; Nelson, L.A.; Rappaport, S., 2001, ApJ, 550, 897
   
\bibitem[jose(2016)]{jose} Jos\'e, J. 2016, in {\em Stellar Explosions - Hydrodynamics and Nucleosynthesis}, (Boca
Raton: CRC Press)

\bibitem[jose(2020)]{jose} Jos\'e, J.; Shore, S.N.; Casanova, J.,  2020, A\&A, 634, 5
   
\bibitem[kiplinger(1979)]{kiplinger} Kiplinger, A.L., 1979, ApJ, 234, 997
   
\bibitem[knigge(2011)]{knigge} Knigge, C.; Baraffe, I.; Patterson, J., 2011, ApJSS, 194, 28
  
\bibitem[koester(2010)]{koester} Koester, D., 2010, MmSAI, 81, 921

\bibitem[kovetz(1988)]{kovetz} Kovetz, A.; Prialnik, D.; Shara, M.M., 1988, ApJ, 325, 828

\bibitem[kuin(2020)]{paul} Kuin, N.P.M.; Page, K.L.; Mróz, P.; et al., 2020, MNRAS, 491, 655
   
\bibitem[kuulkers(2006)]{kuulkers}  Kuulkers, E.; Norton, A.; Schwope, A.; et al., 2006, in {\em Compact Stellar X-Ray Sources}, eds. W.H.G. Lewin and M. van der Klis, Cambridge University Press
   
\bibitem[leibowitz(1992)]{leibowitz} Leibowitz, E.M.; Mendelson, H.; Mashal, E.; et al., 1992, ApJ, 385,49

\bibitem[leibowitz(1993)]{leibowitz} Leibowitz, E.M., 1993, ApJ, 411, 29
   
%\bibitem[Lynden-Bell(1969)]{lynden} Lynden-Bell, ????  

\bibitem[modigliani(2010)]{andrea} Modigliani, A.; Goldoni, P.; Royer, F.; et al., 2010, SPIE, 7737E, 28

\bibitem[mason(2008)]{mason} Mason, E.; Howell, S. B.; Barman, T.S.; et al., 2008, A\&A, 490, 279
   
\bibitem[io(2018)]{io} Mason, E.; Shore, S.N.; De Gennaro Aquino, I.; et al., 2018, ApJ, 853, 27
 
\bibitem[io(2020)]{io} Mason, E.; Shore, S.N.; Kuin, N.P.M.; et al., 2020, A\&A, 635, 115
 
\bibitem[moraes(2009)]{moraes} Moraes, M.; Diaz, M., 2009, ApJ, 138, 1541

\bibitem[mukai(2017)]{mukai} Mukai, K., 2017, PASP, 129, f2001
 
 \bibitem[munari(2012)]{munari} Munari, U.; Moretti, S.; 2012, Balt.A., 21, 22

\bibitem[ness(2012)]{ness} Ness, J.-U.; Schaefer, B.E.; Dobrotka, A.; et al., 2012, ApJ, 745, 43

\bibitem[newsham(2014)]{newsham}  Newsham, G.; Starrfield, S.; Timmes, F.X., in {\em Stella Novae: Past and Future Decades}, ASP Conference Series, Vol. 490, 287, edited by  P.A. Woudt and V.A.R.M. Ribeiro

\bibitem[oliveira(2014)]{oliveira} Oliveira, A.S.; Lima, H.J.F.; Steiner, J.E.; et al., 2014, MNRAS, 444, 2692

\bibitem[prialnik(1986)]{prialnik} Prialnik, D., 1986, ApJ, 310, 222

\bibitem[pringle(1981)]{pringle} Pringle, J.E., 1981, ARA\&A, 19, 137

\bibitem[puebla(2007)]{puebla}  Puebla, R.E.; Diaz, M.P.; Hubeny, I., 2007, AJ, 134, 1923 

%\bibitem[rud(2016)]{rudolf} Rudolf, N.; Günther, H. M.; Schneider, P. C., et al., 2016, A\&A, 585, 113

\bibitem[rauch(2018)]{rauch} Rauch, T.; Demleitner, M.; Hoyer, D.; et al., 2018, MNRAS, 475, 3896

%\bibitem[richman(1996)]{richman} Richman, H.R., 1996, 462, 404

\bibitem[rodriguez-gil(2007)]{rodriguez} Rodriguez-Gil, P.; Schmidtobreick, L.; G\"ansicke, B.T., 2007, MNRAS, 374, 1359

\bibitem[brad(2011)]{brad} Schaefer, B.E.; Pagnotta, A.; LaCluyze, A.P.; et al., 2011, ApJ, 742, 113

\bibitem[linda(2018)]{linda} Schmidtobreick, L.; Mason, E.; Howell, S.B;.; et al., 2018, A\&A, 617, 16

\bibitem[selvelli(2013)]{selvelli} Selvelli, P.L.; Gilmozzi, R.; 2013, A\&A, 560, 49

\bibitem[selvelli(2019)]{selvelli} Selvelli, P.L.; Gilmozzi, R.; 2019, A\&A, 622, 186

\bibitem[shara(1986)]{shara} Shara, M.M.; Livio, M.; Moffat, A.F.J.; et al., 1986, ApJ, 311, 163

\bibitem[shore(1997)]{shore} Shore, S.N.; Starrfield, S.; Ake, T.B.; et al., 1997, ApJ, 490, 393

\bibitem[shore(2016)]{shore} Shore, S.N.; Mason, E.; Schwarz, G.J.; et al., 2016, A\&A, 590, 123

   \bibitem[smette(2015)]{smette} Smette, A.; Sana, H.; Noll, S., et al., 2015, A\&A, 576,77

\bibitem[south(2008)]{south} Southworth, J.; G\''ansicke, B.T.; Marsh, T.; et al., 2008, MNRAS, 391, 591

\bibitem[starrfiled(2016)]{starfriled} Starrfield, S.; Iliadis, C.; Hix, W.R., 2016, PASP, 128, 051001
   %\bibitem[stevenson(1994)]{stevenson} Stevenson, C.C., 1994, MNRAS, 267, 904

\bibitem[take(2013)]{takei} Takei, D.; Sakamoto, T.; Drake, J.J., 2013, AJ, 145, 18
   
\bibitem[tho(2001)]{tho} Thoroughgood, T.D.; Dhillon, V.S.; Littlefair, S.P.; et al., 2001, MNRAS, 327, 1323

   \bibitem[torres(2010)]{torres} Torres, G.; Andersen, J.; Giménez, A., 2010,  A\&ARv, 18, 67
   
\bibitem[teb(2009)]{teb} Tremblay, P. -E.; Bergeron, P., 2009, ApJ, 696, 1755

\bibitem[joel(2011)]{joel} Vernet, J.; Dekker, H; D'Odorico, S.; et al., 2011, 536, 105

\bibitem[Waagen(2013)]{Waagen} Waagen, E.O., 2013, AAN, 492

\bibitem[WJ(2012)]{w e j} Wall, J.V.; Jenkins, C.R.,  2012, {\it Practical Statistics for Astronomers}, 2$^{nd}$ edition, Cambridge University Press 
   
\bibitem[warner(1985)]{warner} Warner, B., 1995, {\it Cataclysmic Variable stars}, Cambridge Astrphysics Series 28, Cambridge University Press
   
\bibitem[yaron(2005)]{yaron} Yaron, O.; Prialnik, D.; Shara, M.M.; et al., 2005, ApJ, 623, 398
   
\end{thebibliography}

\begin{appendix}
\section{Log of observations}
\begin{table*}[h]
\caption{XShooter log of observations.}             % title of Table
\label{table:xsh_obs}      % is used to refer this table in the text
%\centering                          % used for centering table
\scriptsize
\begin{tabular}{c c c c c c c}        % centered columns (4 columns)
\hline\hline                 % inserts double horizontal lines
& UT date & MJD start & exptime & N. of exposures & slit width & readout \\    % table heading 
&  &  & (s) &  & ('') &  \\ \hline  
UVB& 2019-04-14 & 58587.008736778 & 420 & 45 & 1.0 & 1$\times$1,100k/1pt/hg \\   
 &  &  & 400 & 4 &  &  \\   
 &  &  & 380 & 1 &  &  \\   
 &  &  & 350 & 2 &  &  \\   
 &  &  & 150 & 2 &  &  \\   
VIS& 2019-04-14 & 58587.008796968 & 420 & 54 & 0.9  & 1$\times$1,100k/1pt/hg \\   
NIR$\dagger$& 2019-04-14 & 58587.00883251 & 189$\times$3 & 11 & 0.6 & NonDest \\ 
&  &  & 180$\times$3 & 43 &  &  \\ \hline  
UVB& 2019-04-27 & 58600.003039113 & 400 & 56 & 1.0 & 1$\times$1,100k/1pt/hg \\  
VIS& 2019-04-27 & 58600.003099419 & 400 & 56 & 0.9 & 1$\times$1,100k/1pt/hg \\   
NIR$\dagger$& 2019-04-27 & 58600.00313732 & 180$\times$3 & 56 & 0.6 & NonDest \\ 
\hline                                   %inserts single line
\end{tabular}
\\ $\dagger$ For the NIR spectrograph the exposure time is given in DIT$\times$NDIT. Note also that the readout mode is fixed in the NIR. 
\end{table*}
\begin{table}[h]
\caption{Swift log of observations}             % title of Table
\label{table:log_uv}      % is used to refer this table in the text
%\centering                          % used for centering table
\scriptsize
\begin{tabular}{c c c c c c}        % centered columns (4 columns)
\hline\hline                 % inserts double horizontal lines
 UT date & MJD start & exptime & filter~\ddag &  mag & flux \\    % table heading 
         &           & (s)     &        &   (Vega)   & $\times$10$^{-14}$~erg/s/cm$^2$/\AA \\\hline  
 2019-04-21 & 58594.98145 & 161.4 &  UVM2 & 12.47$\pm$0.04 & 4.77$\pm$0.12    \\ 
 2019-04-22 & 58595.31350 & 151.9 &  UVW1 & 12.47$\pm$0.04 & 4.09$\pm$0.16  \\      % inserting body of the table
 2019-04-22 & 58595.32361 & 60.3  &  UVW2 & 12.19$\pm$0.04 & 7.10$\pm$0.24  \\  
 2019-06-19 &  58653.41332 & 403.7 &  U   & 12.46$\pm$0.03 & 3.66$\pm$0.10 \\  
 2019-06-19 &  58653.41834 & 32.3  &  UVW1 & 12.39$\pm$0.04 & 4.41$\pm$0.20  \\  
 2019-06-19 &  58653.42264 & 24.9  &  UVW2 & 12.25$\pm$0.05 & 6.71$\pm$0.27  \\  
 2019-06-19 &  58653.42319 & 32.3  &  UVW1 & 12.42$\pm$0.04 & 4.28$\pm$0.18  \\  
 2019-06-19 &  58653.42749 & 25.3  &  UVW2 & 12.22$\pm$0.05 & 6.92$\pm$0.27  \\  
 2019-06-19 &  58653.88337 & 37.2  &  UVW1 & 12.32$\pm$0.04 & 4.68$\pm$0.18  \\  
 2019-06-19 &  58653.88778 & 37.2  &  UVW1 & 12.38$\pm$0.04 & 4.45$\pm$0.19  \\  
 2019-06-19 &  58653.89272 & 29.1  &  UVW2 & 12.21$\pm$0.05 & 6.98$\pm$0.27  \\  
\hline                                   %inserts single line
\end{tabular}
\ddag \ Filters nominal wavelength and width:\\
u: 3501, 785 \AA \\
uvw1: 2591, 693 \AA \\
uvm2: 2229, 498 \AA \\
uvw2: 2033, 653 \AA
\end{table}
\begin{table}[h]
\caption{Chandra log of observations}             % title of Table
\label{table:x_obs}      % is used to refer this table in the text
%\centering                          % used for centering table
\scriptsize
\begin{tabular}{c c c c c}        % centered columns (4 columns)
\hline\hline                 % inserts double horizontal lines
Obs. ID & UT date & MJD start & exptime & rate \\    % table heading 
       &            &  & (ks)  & ($\times$10$^{-4}$ cts/s) \\ \hline  
 22682 & 2019-09-02 & 58728.10625 & 14.87 & 141.3$\pm$9.92\\      % inserting body of the table
\hline                                   %inserts single line
\end{tabular}
\end{table}

\section{Emission line list}\label{sec:linelist} 
In this appendix we list the emission lines observed in the XShooter spectra. A few weak lines could not be identified and have a question mark in the "line ID" column. Uncertain identification show the ion followed by a question mark in the same column. 
In the "formation site" column, we indicate an ejecta or binary origin with the syllable "ej" and "bi", respectively. We indicate as "ej" only, all the emission lines whose trailed spectrogram did not show moving features. However, for some of them there must but a binary contribution as well, if other transitions of the same series show "moving features". This is the case for example of H$_\beta$ and other H emission lines. The lack of visible binary contribution is explained by different optical depth (see Section \ref{sec:emlinespc}).

\begin{table}[h]
\caption{Emission lines in the UVB arm of the XShooter spectrum}             % title of Table
\label{table:uvb_lines}      % is used to refer this table in the text
%\centering                          % used for centering table
\scriptsize
\begin{tabular}{c c c p{2cm}}        % centered columns (4 columns)
\hline\hline                 % inserts double horizontal lines
line ID & rest wavelength & formation site &  note\\    % table heading 
        &                 &                &  \\ \hline  
O~III & 3132.9 & ej + bi & weak bi \\      % inserting body of the table
? & 3148 & ej & weak \\  
He~I & 3187.7 & ej & \\  
He~II & 3203.2 & ej + bi & weak bi \\  
O~III & 3265.5 & ej & \\  
$[$Ne~V$]$ & 3345.9 & ej & \\  
O~III & 3408.1 & ej & \\  
O~III & 3444.1 & ej & \\  
O~III & 3702.8+3703.4 & ej & blend, + H$_{16}$\\  
O~III & 3712.5 & ej & blend, + H$_{15}$\\  
O~III & 3725.3 & ej & blending with previous 2\\  
O~III & 3759.9 & ej & blend\\  
H$_{10}$ & 3797.8 & ej+bi & strong bi\\  
He~I? & 3819.9 & bi & very weak\\  
H$_\eta$ & 3835.4 & ej+bi & strong bi \\  
$[$Ne~III$]$ & 3868.7 & ej & \\  
H$_\zeta$ & 3889.2 & ej+bi & \\
H$_\varepsilon$ & 3970.1 & ej+bi & blend with $[$Ne~III$]$\\  
He~I & 4026.2 & ej+bi & \\  
C~III? & 4070.3 & ej & weak \\  
H$_\delta$ & 4101.7 & ej+bi & blending with pure ej lines \\  
He~I & 4143.8 & ej+bi & \\  
N~III & $\sim$4198 & ej & big blend\\  
O~III & 4239.5 & ej & very weak \\  
C~II & 4267.0 & ej+bi & \\  
H$_\gamma$ & 4340.5 & ej+bi & \\  
$[$O~III$]$ & 4363.2 & ej & \\  
N~III? & 4379.1 & ej & \\  
He~I & 4471.3 & ej+bi & mostly bi\\  
N~III? & 4510.9 & ej & \\
N~III & 4523.6 & ej & \\  
He~II & 4541.6 & ej & \\  
N~II? & 4607.2 & ej & or $[$Ar~V$]$?\\  
N~III & 4640.6+4641.9 & ej & +N~III~4634.2 +C~III 4650.2 \\  
He~II & 4685.7 & ej+bi & \\  
H$_\beta$ & 4861.3 & ej & +bi \\  
$[$O~III$]$ & 4959 & ej & \\  
$[$O~III$]$ & 5007 & ej & saturated\\  
? & $\sim$5037 & ej & \\  
? & $\sim$5056 & ej & \\  
$[$Fe~VI$]$ & 5146.8 & ej & \\  
? & $\sim$5197 & ej & \\  
He~II & 5411 & ej+bi & \\  
\hline                                   %inserts single line
\end{tabular}
\end{table}

\begin{table}[h]
\caption{Emission lines in the VIS arm of the XShooter spectrum}             % title of Table
\label{table:vis_lines}      % is used to refer this table in the text
%\centering                          % used for centering table
\scriptsize
\begin{tabular}{c c c p{2cm}}        % centered columns (4 columns)
\hline\hline                 % inserts double horizontal lines
line ID & rest wavelength & formation site & note\\    % table heading 
        &                 &                & \\ \hline  
$[$Fe~VI$]$ &  5678.0 & ej & \\  
$[$N~II$]$ & 5754.8 & ej & \\  
C~IV & 5801.5 & ej+bi & \\  
He~I & 5875.8 & ej+bi & \\  
$[$O~I$]$ & 6300.2 & ej & \\  
$[$O~I$]$ & 6363.9 & ej & \\  
H$_\alpha$+$[$N~II$]$ & 6562.8 & ej & blend, saturated\\  
He~I & 6678.1 & ej+bi & \\  
$[$S~II$]$? & 6731.3 & ej & \\  
He~I & 7065.2 & ej+bi & \\  
$[$Ar~III$]$ & 7135.8 & ej & blending \\  
$[$Ar~IV$]$ & 7236.0 & ej & +telluric \\  
$[$Ar~IV$]$ & 7332.0 & ej & blending \\  
? & $\sim$7731 & ej & \\  
O~I & 7773 & ej+bi & \\  
C~III? & 8021 & ej & \\  
He~II? & 8236.8 & ej & +telluric\\  
Pa$_{27}$ & 8306.1 & ej & +telluric\\  
O~I+Pa$_{18}$ & 8446+8438 & ej+bi & blend\\  
Pa$_{16}$ & 8502.5 & ej+bi & \\  
Pa$_{15}$ & 8545.4 & ej+bi & \\  
Pa$_{14}$ & 8598.4 & ej+bi & \\  
Pa$_{13}$ & 8665.6 & ej+bi & \\  
Pa$_{12}$ & 8750.5 & ej+bi & \\  
Pa$_{11}$ & 8862.8 & ej+bi & \\  
Pa$_{10}$ & 9014.9 & ej+bi & \\  
$[$S~III$]$ & 9069.4 & ej & +telluric\\  
Pa$_9$ & 9229.0 & ej+bi & +telluric\\  
$[$S~III$]$ & 9532.1 & ej & +Pa$_8$ 9546 +telluric\\  
Pa$_7$ & 10049.4 & ej+bi & \\  
He~II & 10123.6 & ej+bi & \\  
\hline                                   %inserts single line
\end{tabular}
\end{table}

\begin{table}[h]
\caption{Emission lines in the NIR arm of the XShooter spectrum. All wavelengths are in air.}             % title of Table
\label{table:nir_lines}      % is used to refer this table in the text
%\centering                          % used for centering table
\scriptsize
\begin{tabular}{c c c p{2cm}}        % centered columns (4 columns)
\hline\hline                 % inserts double horizontal lines
line ID & rest wavelength & formation site & note\\    % table heading 
        &                 &                &  \\ \hline  
Pa$_7$ & 10049.4 & ej+bi & \\  
He~II & 10123.6 & ej+bi & \\  
Pa$_6$ & 10938.1 & ej+bi & \\  
Pa$_5$ & 12818.1 & ej & \\  
Br$_{17}$ & 15438.9 & ej+bi & \\  
Br$_{16}$ & 15556.4 & ej+bi & +sky\\  
Br$_{15}$ & 15700.7 & ej+bi & +telluric\\  
Br$_{14}$ & 15880.5 & ej+bi & \\  
Br$_{13}$ & 16109.3 & ej+bi & \\  
Br$_{12}$ & 16407.2 & ej+bi & blending wt 16436.5 (ej)\\  
Br$_{11}$ & 16806.5 & ej+bi & \\  
He~I & 17002.5 & ej+bi & almost no ej\\  
Br$_{10}$ & 17362.1 & ej+bi & +sky\\
Br$_{7}$ & 21655.3 & ej+bi & very weak bi\\  
? & 21145 & ej & likely a blend\\
? & 21874 & ej & weak\\  
\hline                                   %inserts single line
\end{tabular}
\end{table}
\end{appendix}

\end{document}